\documentclass[journal,twocolumn]{IEEEtran}
\usepackage[normalem]{ulem}
\usepackage{blindtext, graphicx}
\usepackage{listings}
\usepackage{graphicx}
\usepackage[font=small]{caption}
\usepackage{color}
\usepackage{amsmath,bm}
\usepackage{amsmath}
\usepackage{amssymb}
\usepackage{algorithm}
\usepackage{algpseudocode}
\usepackage{amsthm}
\usepackage{makecell}

\newtheorem{proposition}{Proposition}
\usepackage[utf8]{inputenc}

\makeatletter
\let\NAT@parse\undefined
\makeatother
\usepackage{hyperref}

\title{Optimal Vehicle Charging in Bilevel Power-Traffic Networks via Charging Demand Function}

\author{Yufan Zhang, Sujit Dey~\IEEEmembership{Fellow,~IEEE,} Yuanyuan Shi~\IEEEmembership{Member,~IEEE.}
\vspace{-2em}
\thanks{The authors are with the Department of Electrical and Computer Engineering, University of California San Diego, San Diego, California 92093, US.}
}

\begin{document}

\maketitle
\thispagestyle{empty}
\pagestyle{plain}

\begin{abstract}
Electric vehicle (EV) charging couples the operation of power and traffic networks. Specifically, 
the power network determines the charging price at various locations, while EVs on the traffic network optimize the charging power given the price, acting as price-takers.
We model such decision-making processes by a bilevel program,
with the power network at the upper-level and the traffic network at the lower-level. However, since the two networks are managed by separate entities and the charging expense term, calculated as the product of charging price and charging demand, is nonlinear. Solving the bilevel program is nontrivial. To overcome these challenges, we derive the charging demand function using multiparametric programming theory. This function establishes a piecewise linear relationship between the charging price and the optimal charging power, enabling the power network operator to manage EV charging power independently while accounting for the coupling between the two networks. With the derived function, we are also able to replace the nonlinear charging expense term with a piecewise quadratic one, thus guaranteeing solution optimality. Our numerical studies demonstrate that different traffic demands can have an impact on charging patterns and the power network can effectively incentivize charging at low-price nodes through price setting.

Keywords: electric vehicles, bilevel program, coupled networks, charging demand function

\end{abstract}

\section{Introduction}

There is a growing trend towards electrifying a significant portion of the transportation sector to decarbonize \cite{XIE20211908}. As of 2022, electric vehicle (EV) sales accounted for 10\% of all new car sales globally \cite{EV2022}. The increasing number of EVs has raised concerns about their impact on the power network \cite{muratori2018impact}. Without proper coordination, EV charging can compromise grid reliability. As reported in \cite{ELHELOU2022100093}, as little as 11\% of heavy-duty vehicles charging simultaneously in Texas can lead to significant voltage violations on the transmission network. Therefore, careful management of EV charging is crucial.

Modeling a charging station as an aggregator is one of the approaches for organizing charging smartly \cite{xu2014challenges,yi2021aggregate}. 
This approach treats the charging demand at a station as a flexible load that can be coordinated to meet the needs of the grid. For example, it allows for EV charging to be delayed until there is enough generation capacity on the grid, preventing high charging demand from compromising grid reliability \cite{lacey2017smart}.
Previous studies have investigated the use of aggregated EV charging for various purposes, such as frequency regulation \cite{lauinger2020reliable}, ramping provision \cite{6582687}, and congestion relief \cite{li2013distribution}, to name a few. To achieve these goals, these studies treat EV driving behavior information such as arrival time, charging duration, and energy demand as input parameters \cite{yi2021aggregate}, which can be obtained by forecasting approaches. From the perspective of aggregators, the traveling plan of EVs and the charging management are viewed as two separate decision-making processes.
Although this assumption can simplify the analysis, the coupling of those decision-makings is ignored, which may lead to some unreliable decisions \cite{Optpricing}. 
For example, if the EV charging manager fails to consider the fact that EVs may travel to other charging stations in response to electricity prices, it could result in a charging demand hotspot where many EVs travel to a single station with low electricity prices, overloading the charging station.


Linking vehicle travel plans with charging management has been advocated in recent years \cite{chen2018two}. For optimizing the collective driving behavior, an entity represented by an independent traffic system operator (ITSO), is introduced for settling the travel plans of vehicle drivers. 
Specifically, ITSO solves the optimal traffic assignment problem, determining routes, charging locations, and charging power for all vehicles in a way that minimizes total traveling and charging costs. ITSO acts as a price-taker to the locational marginal price (LMP) of electricity, which is determined by the optimal power flow (OPF) problem solved by the independent distribution system operator (IDSO).
Research has investigated the joint optimization of ITSO's traffic assignment problem and IDSO's OPF problem, demonstrating that the social welfare of the coupled networks can be achieved under the LMP mechanism \cite{Optpricing,8737720}. Also, other studies have taken a game-theoretic perspective, with both ITSO and IDSO seeking to optimize their own costs \cite{ZHOU2021116703}. These studies have shown that the power-traffic system game is a potential game, with the Nash equilibrium coinciding with the joint optimization solution.

One challenge of coordinating the coupled power and traffic networks using the joint optimization framework is that the two network operators belong to separate entities, making it difficult to disclose the entire decision model to each other. Recent works have resorted to distributed optimization for solving the joint optimization problem, requiring multiple rounds of information exchange between ITSO and IDSO \cite{Optpricing,8737720,ZHOU2021116703}. However, such iterative updates are not currently within the power network operation structure, and iterative algorithms may not meet the time requirements of the power network operation, as the lack of two-way real-time communication between the systems makes multiple rounds of information exchange time-consuming \cite{ZHENG2022446}.

To maintain the existing operation framework of the power network, it is desirable to solve the OPF problem independently and efficiently while accounting for the coupling between power and traffic networks. One possible approach is to model the interaction between the two networks as a bilevel program \cite{BilevelReview}, with the power network at the upper-level determining the LMP and the traffic network at the lower-level determining the optimal charging demand given the price.
Interestingly, this bilevel formulation is equivalent to the joint optimization and game-theoretic formulations proposed in previous studies \cite{Optpricing,8737720,ZHOU2021116703}. We will explore this equivalence further in Section \ref{sec:equivalence}.

However, solving the bilevel power-traffic optimization problem is not trivial. Firstly, the charging expense that needs to be minimized is a product of the charging price (an upper-level decision variable) and the charging demand (a lower-level decision variable). The resulting bilinear term introduces nonlinearity into the problem, making it difficult to solve with an optimality guarantee. Secondly, solving the bilevel problem requires a collection of the Karush-Kuhn-Tucker (KKT) conditions of the lower-level problem, which in turn requires detailed knowledge of the ITSO decision model.

To address these challenges, we propose a novel approach for solving the bilevel power-traffic optimization problem. Our approach is based on the observation that the decision problems of the power and traffic networks are coupled by the charging price and demand. If the relationship between the charging price and the optimal charging demand can be obtained, the IDSO can use such a function for solving the OPF problem independently, without the need of knowing the detailed decision model of ITSO. Similar to the concept of demand function~\cite{kirschen2018fundamentals} (which expresses the demand as a function of price), we define such a relationship as the \emph{charging demand function}.

As a critical step in our approach, we derive the charging demand function using multiparametric programming theory \cite{tondel2003algorithm, Grancharova2012}. This function describes the relationship between the charging price and the optimal charging power, which is determined by the ITSO. By introducing the charging demand function, we are able to ``kill two birds with one stone''. On the one hand, the power network can set the charging price and manage EV charging power by solving the OPF problem independently, which aligns with the current operation framework of the power network. 
Furthermore, the charging demand function is proven to be piecewise linear, allowing us to transform the bilinear term in the charging expense into a piecewise quadratic term. This transformation makes the bilevel program for solving EV charging management with \emph{optimality guarantees}.

Compared with existing studies, the main contributions of the paper are:




1) A theoretical derivation of the charging demand function, which describes the relationship between the charging price and the optimal charging demand as a piecewise linear function. This function allows for efficient vehicle travel and charging management by the power network itself.

2) A solution strategy to the bilevel program that replaces the lower-level ITSO problem with the charging demand function. This transformation converts the bilinear term of the charging expense into a piecewise quadratic term, enabling a computationally efficient solution of the bilevel power-traffic program with optimality guarantees.

3) A complimentary understanding of different perspectives of modeling the power and traffic networks interaction. We show our bilevel program is equivalent to the models from the perspectives of joint optimization and game theory, but using different solution techniques that do not require detailed traffic models nor iterative information exchange. 

The remaining parts of this paper are organized as follows. Section \uppercase\expandafter{\romannumeral2} presents the bilevel setup and the decision-making problems of power and traffic networks as preliminaries, whereas the details of charging demand function derivation and solution strategy are given in Section \uppercase\expandafter{\romannumeral3}. Results are discussed and evaluated in Section \uppercase\expandafter{\romannumeral4}, followed by the conclusion and future works.

\textit{Notation:} The notation $\bm{x}_{\mathcal{I}}=[x_i]_{i \in \mathcal{I}}$ denotes a column vector comprised by elements in the set $\mathcal{I}$. $|\cdot|$ denotes the cardinality of a set. $\bm{I}_M$ denotes an identity matrix with the size of $\mathbb{R}^{M \times M}$. $\bm{1}_M$ is an all-one column vector with the dimension of $\mathbb{R}^M$. $\bm{O}$ is an all-zero matrix with a size to be defined by the use case. Given the column vectors $\bm{x}_1,\bm{x}_2$, we use the notation $[\bm{x}_1;\bm{x}_2]$ to denote the vertical stack of them.

\section{Preliminaries: Bilevel Setup and Networks Model}

In this section, we first introduce the interaction between the power and traffic networks as a bilevel framework in Section~\ref{Bilevel Setup}. The decision problems of traffic and distribution networks are described in Section~\ref{traffic Network Model} and Section~\ref{Distribution Network Model}, respectively.

\subsection{Bilevel Setup}\label{Bilevel Setup}
\begin{figure}[hbtp]
  \centering
  \includegraphics[scale=0.5]{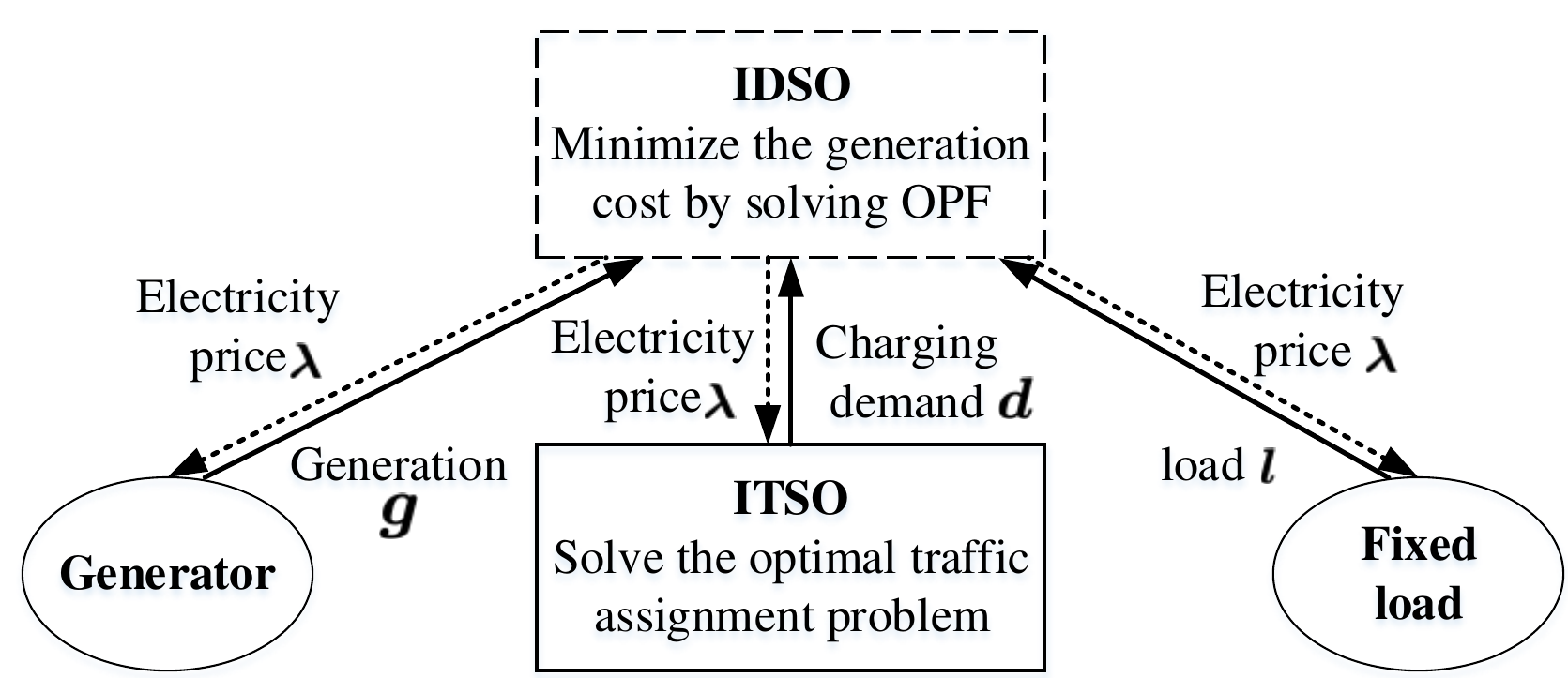}\\
  \caption{Diagram of the bilevel setup. The problem of the ITSO is at the lower-level, while the problem of IDSO is at the upper-level. 
  }\label{Fig 1}
\end{figure}
The setup of the problem is sketched in Fig. \ref{Fig 1}. We study the system-level pricing strategy of non-profit IDSO, who manages the distribution network and solves the OPF problem to minimize the generation cost for satisfying the fixed load and EV charging demand. The resulting nodal electricity price, namely LMP, affects the EVs’ routing and charging decisions. The traffic network is managed by the non-profit ITSO, who optimally assigns the traffic flow to minimize the travel cost brought by congestion and charging expenses. The power and traffic networks are coupled by common infrastructure, namely the EV charging stations. We follow the common assumption that the ITSO acts as a price-taker in the decision problem, which optimizes EVs' charging behaviors given the charging price determined by IDSO. 
To model such a relationship, a bilevel framework is considered here, where the decision problem of IDSO solving the OPF is at the upper-level, and the optimal traffic assignment problem of ITSO is at the lower-level. The two problems are coupled by \textcolor{black}{the decisions regarding the electricity price $\lambda_i$ and the corresponding optimal charging demand $d_i^*$ at the node $i$ connected to the charging station.} 
The relationship between $\lambda_i$ and $d_i^*$ will be discussed in detail in Section \uppercase\expandafter{\romannumeral3}. For simplicity, we consider a single-period problem but the formulation can be extended to the multi-period case as future work.

\subsection{Traffic Network Model}
\label{traffic Network Model}

The traffic network is modeled by a connected directional graph $\mathcal{R}=(\mathcal{V},\mathcal{A})$, where $\mathcal{V}$ and $\mathcal{A}$ denote the sets of vertexes and arcs. 
We model the traveling time on arc $a$ as a linear function $\tau_a(\xi_a)$ such that the traveling time increases with the increase of the traffic flow $\xi_a$ on arc $a \in \mathcal{A}$. ; see \cite{Optpricing}.
\begin{align} &\tau_a(\xi_a)=\xi_a^0+\xi_a / R_a,\forall a \in \mathcal{A}\label{1(a)}
\end{align}
where $\xi_a^0, R_a$ are the parameters related to the shape of the traveling time function.

\begin{figure}
  \centering
  \includegraphics[scale=0.5]{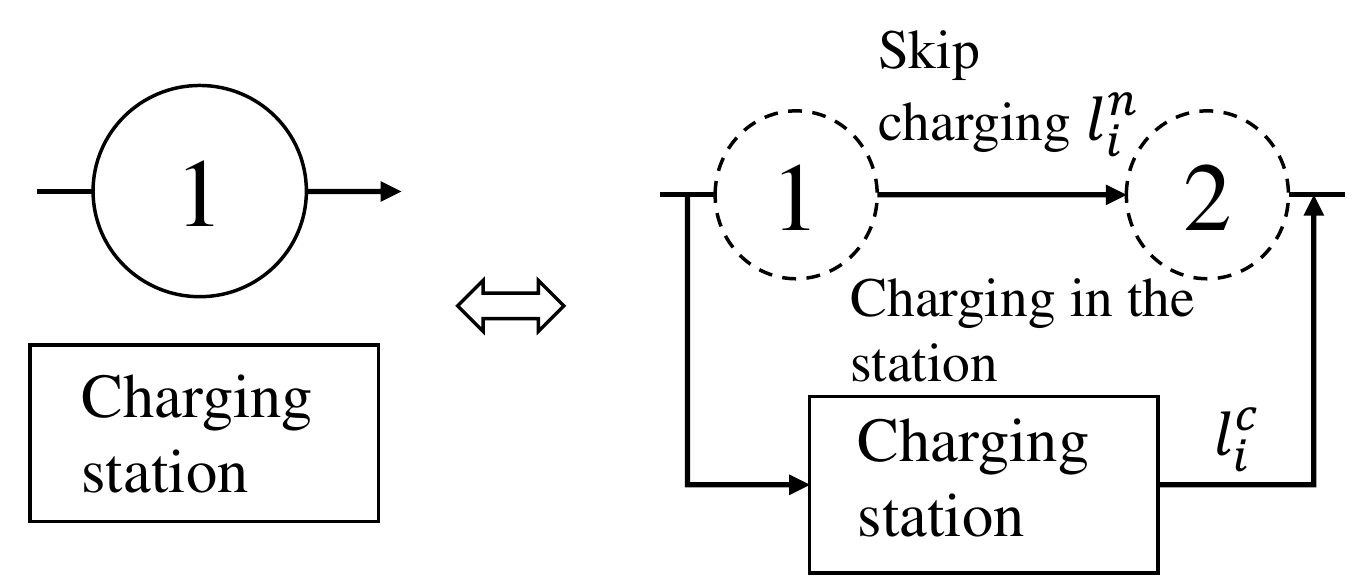}\\
  \caption{The illustration of virtual arcs to represent the charging behavior in a charging station. An extra node (node \#2) is introduced to form the virtual arcs. 
  }
  \label{Fig 2}
\end{figure}

Let $\mathcal{C}\in \mathcal{V}$ denote the set of nodes where the charging stations are located in. We model the nodes of charging stations as ``virtual arcs'' following \cite{Optpricing} (see Fig. \ref{Fig 2}). At a node $i \in \mathcal{C}$, the EV driver can choose the arc $l_i^c$ for charging or the arc $l_i^n$ for skipping charging. The former decision incurs costs of time and electricity, while the latter one incurs no cost. Concretely,  
the time costs for an EV on a charging arc $l_i^c$ consist of two parts: waiting time and charging time. 
The waiting time increases as the flow of EVs $\xi_a,\forall a \in l_i^c$ increases and is modeled using the function in \eqref{1(a)} \cite{davidson1966flow} as well. 
The charging time is associated with the average charging demand $e_a$, which is given by $e_a/\rho_a$, where $\rho_a$ is the charging rate. The sum of waiting and charging time
constitutes the latency time spent in a charging station, i.e., 
\begin{equation} \label{2}
    \tau_a(\xi_a)=e_a/\rho_a + \xi_a^0+\xi_a / R_a ,\forall a \in \cup_{i \in \mathcal{C}}l_i^c
\end{equation}

For an EV on the virtual arc $l_i^n$, the time cost is zero, i.e.,

\begin{equation}\label{3}
\tau_a(\xi_a)=0,\forall a \in \cup_{i \in \mathcal{C}}l_i^n  
\end{equation}

With the notion of the virtual arcs, the extended traffic network is introduced, with the set of arcs given by
\begin{equation}\label{4}
    \mathcal{A}^e=\cup_{i \in \mathcal{C}} (l_i^c \cup l_i^n) \cup \mathcal{A}
\end{equation}



The ITSO optimizes the aggregated 
cost of total traffic system latency time and charging expense 
which is given by
\begin{equation}\label{5}
    \sum_{a \in \mathcal{A}^e} \gamma \cdot \xi_a \cdot \tau_a(\xi_a)  + \sum_{a \in \cup_{i \in \mathcal{C}}l_i^c} \lambda_i \cdot e_a \cdot \xi_a \,,
\end{equation}
where $\gamma$ is the cost coefficient for one unit of travel time, $\lambda_i$ is the electricity price at the charging station $i \in \mathcal{C}$, which is obtained as the optimal dual solution of the OPF problem solved by IDSO. 
From now on, we use the notation $d_i$ to denote the charging demand in charging station $i$,
\begin{equation}\label{6}
d_i=e_a \cdot \xi_a,\forall a \in \cup_{i \in \mathcal{C}}l_i^c 
\end{equation}


 Let the set $\mathcal{W}$ denote the set of finite origin–destination (O–D) pairs. $\forall w \in \mathcal{W}$, there are a set of feasible routes $\mathcal{R}_w$ that allows the driver to finish the trip on the extended graph. Let $m_w$ be the travel demand of O-D pair $w$ and $f_w^k$ be the flow on route $k \in \mathcal{R}_w$. We use $\bm{f}_w \in \mathbb{R}^{|\mathcal{R}_w|}=[f_w^k]_{k \in \mathcal{R}_w}$ to denote the vector formed by the route flows of the O-D pair $w$, and define $\bm{f} = [\bm{f}_w]_{w \in \mathcal{W}}$ as a collection of all the $\bm{f}_w$. 
 Let $\bm{\xi}=[[\xi_a]_{a \in \mathcal{A}};[\xi_a]_{a \in \cup_{i \in \mathcal{C}}l_i^n};[\xi_a]_{a \in \cup_{i \in \mathcal{C}}l_i^c}]$ denote the arc flows on $\mathcal{A}^e$. 
The dimension of $\bm{\xi}$ is $|\mathcal{A}^e|$, which is equal to $|\mathcal{A}|+2\cdot|\mathcal{C}|$. To link the arc with the route, the link matrix $\bm{A}_w \in \mathbb{R}^{|\mathcal{A}^e| \times |\mathcal{R}_w|}$ is introduced for each O-D pair $w \in \mathcal{W}$, where the element in the $a_{th}$ row is assigned with 1 if the arc $a$ is on the route belonging to $\mathcal{R}_w$ and 0 otherwise. With these notations, the decision problem of ITSO is formulated as
\begin{subequations}\label{7}
\begin{align} &\mathop{\min}_{\bm{\xi},\bm{f}}\sum_{a \in \mathcal{A}^e} \gamma \cdot \xi_a \cdot \tau_a(\xi_a) + \sum_{i \in \mathcal{C}} \lambda_i \cdot d_i 
\label{7(a)}\\ 
&s.t. \ \  \eqref{6} \label{7(b)}\\
& \qquad \bm{1}_{|\mathcal{R}_w|}^\top \cdot \bm{f}_w=m_w, \forall w \in \mathcal{W}
\label{7(c)}\\
& \qquad  \bm{\xi}=\sum_{w \in \mathcal{W}} \bm{A}_w \cdot \bm{f}_w \label{7(d)}\\
& \qquad  0 \leq \bm{\xi} \leq \bm{\overline{\xi}}\label{7(e)}
\end{align}
\end{subequations}
where \eqref{7(c)} ensures that the sum of traffic flow on the routes of the O-D pair $w$ equals the travel demand $m_w$. 
\eqref{7(d)} maps the flow on routes to the flow on arcs. \eqref{7(e)} ensures the traffic flows on the arcs  are non-negative and within the limits.  With the quadratic term introduced by the first term in \eqref{7(a)}, the decision problem of ITSO is a quadratic program (QP).

\textbf{Remark 1:} We note that the decision problem of ITSO in \eqref{7} can be extended to the case considering EV discharging. For that, the battery degradation cost $c(d_i)$ will be considered in the objective, and modeled by a quadratic function \cite{bian2023predicting}. Also, the charging/discharging demand is modeled in a range, i.e., $-e_a^d \cdot \xi_a \leq d_i \leq e_a^c \cdot \xi_a,\forall a \in \cup_{i \in \mathcal{C}}l_i^c$, where average discharging/charging rates $e_a^d,e_a^c$ are constants. When all the EVs in the traffic flow of arc $a$ charge, the charging demand $d_i$ equals $e_a^c\cdot\xi_a$. When all the EVs discharge, the charging demand $d_i$ equals $-e_a^d \cdot \xi_a$. When part of the EVs charge and the remaining ones discharge, the charging demand is in the middle of the range. To sum up, considering discharging, the traffic network model is still a QP.

\subsection{Distribution Network Model}\label{Distribution Network Model}

We use a graph $\mathcal{G}(\mathcal{N},\mathcal{L})$ to represent the power distribution network, where $\mathcal{N}$, $\mathcal{L}$ are the sets of nodes and lines, respectively. As charging stations are the common infrastructures of distribution and traffic systems, we use $\mathcal{C}$ to denote the charging stations. 
To satisfy the charging demand and fixed load, IDSO optimally dispatches the generators under the network constraint and solves the OPF problem. Let $\bm{g}=[g_i]_{i \in \mathcal{N}}$, $\bm{v}=[v_i]_{i \in \mathcal{N}}$, $\bm{\theta}=[\theta_i]_{i \in \mathcal{N}}$ and $\bm{l}=[l_i]_{i \in \mathcal{N}}$ denote the vectors of generation, the voltage amplitude and phase angle, and fixed load, respectively, and $\bm{d}=[d_i]_{i \in \mathcal{N}}$ denote the vector of charging demand, where $d_i=0,\forall i \in \mathcal{N}\backslash \mathcal{C}$.
Given the generation cost coefficient $\bm{c}=[c_i]_{i \in \mathcal{N}}$, the problem of IDSO minimizes the total generation cost subject to the demand requirement and power flow constraints,
\begin{subequations}\label{8}
\begin{align}
&\mathop{\min}_{\bm{g},\bm{v},\bm{\theta}}\boldsymbol{c}^\top\boldsymbol{g}\label{8(a)}\\
&s.t. \ \ 
0 \leq \boldsymbol{g} \leq \overline{\boldsymbol{g}}:\underline{\boldsymbol{\tau}},\overline{\boldsymbol{\tau}}\label{8(b)}\\
& \quad \ l_i+d_i + \sum_{j \in \Omega_i}K_{1ij}(v_i-v_j)+\nonumber \\
& \quad  \quad  \quad\sum_{j \in \Omega_i}K_{2ij}(\theta_i-\theta_j) =g_i:\lambda_i,\forall i \in \mathcal{N} \label{8(c)}\\
& \quad \ 
-f_{ij}\leq K_{1ij}(v_i-v_j)+K_{2ij}(\theta_i-\theta_j) \leq f_{ij}:\nonumber \\
& \qquad 
\qquad \qquad \qquad \quad \underline{\eta_{ij}},\overline{\eta_{ij}},\forall i \in \mathcal{N},\forall j \in \Omega_i \label{8(d)}\\
& \quad \
\underline{v}\leq v_i \leq \overline{v}:  \underline{\mu_i},\overline{\mu_i},\forall i \in \mathcal{N},
\label{8(e)}
\end{align}
\end{subequations}
where the dual variables are given after colons, and $\underline{\bm{\tau}}=[\underline{\tau_i}]_{i \in \mathcal{N}},\overline{\bm{\tau}}=[\overline{\tau_i}]_{i \in \mathcal{N}}$. Under the generation capacity $\overline{\boldsymbol{g}}$, \eqref{8(b)} limits the range of generation. Let $\Omega_i$ denote the set of buses connected to node $i$. \eqref{8(c)} is the nodal power balance constraint and linearized
distribution power flow model \cite{AMulti} is adopted, where $K_{1ij}=\frac{x_{ij}r_{ij}}{r_{ij}^2+x_{ij}^2},K_{2ij}=\frac{x_{ij}^2}{r_{ij}^2+x_{ij}^2}$ are the parameters calculated by the line resistance $r_{ij}$ and line reactance $x_{ij}$. \eqref{8(d)} and \eqref{8(e)} bound the range of power flow and voltage amplitude, with $f_{ij}$ and $\underline{v},\overline{v}$ as the limits. 

For the IDSO decision problem in \eqref{8}, the charging demand $d_i$ serves as the parameters and the electricity price $\lambda_i$ is the dual variable of the constraint \eqref{8(c)}. 
To clearly show the relationship between the problems of IDSO and ITSO, the dual problem of \eqref{8} is derived, whose derivation is achieved by restricting the derivative of Lagrange function of \eqref{8} to $\theta_i,v_i,g_i$ to zero. 
We define the vectors $\bm{\overline{\mu}}=[\overline{\mu_i}]_{i \in \mathcal{N}},\bm{\underline{\mu}}=[\underline{\mu_i}]_{i \in \mathcal{N}},\bm{\lambda}=[\lambda_i]_{i \in \mathcal{N}},\bm{\underline{\eta}}=[\underline{\eta_{ij}}]_{i \in \mathcal{N},j \in \Omega_i},\bm{\overline{\eta}}=[\overline{\eta_{ij}}]_{i \in \mathcal{N},j \in \Omega_i}$ as the collections of dual variables. The dual problem of \eqref{8} is given by, 
\begin{subequations}\label{9}
\begin{align}
&\mathop{\max}_{\bm{\overline{\tau}},\bm{\underline{\tau}},\bm{\overline{\mu}},\bm{\underline{\mu}},\bm{\lambda},\bm{\underline{\eta}},\bm{\overline{\eta}}} \sum_{i \in \mathcal{N}}-\overline{\mu_i}\cdot\overline{v_i}+\underline{\mu_i}\cdot\underline{v_i}-\overline{\tau_i}\cdot\overline{g_i}- \nonumber \\
& \sum_{i\in\mathcal{N}}\sum_{j \in \Omega_i}(f_{ij}\overline{\eta_{ij}}+f_{ij}\underline{\eta_{ij}})+\sum_{i \in \mathcal{N}}\lambda_i\cdot l_i+\sum_{i \in \mathcal{C}}\lambda_i\cdot d_i \label{9(a)}\\
& s.t. \sum_{j\in\Omega_i}K_{2ij}(\lambda_i-\lambda_j+\overline{\eta_{ij}}-\overline{\eta_{ji}}-\underline{\eta_{ij}}+\underline{\eta_{ji}})=0,\forall i \in \mathcal{N}\label{9(b)}\\
& \quad \ \sum_{j\in\Omega_i}K_{1ij}(\lambda_i-\lambda_j+\overline{\eta_{ij}}-\overline{\eta_{ji}}-\underline{\eta_{ij}}+\underline{\eta_{ji}})+\nonumber \\
& \qquad \qquad \qquad \qquad \qquad \qquad \qquad \overline{\mu_i}-\underline{\mu_i}=0,\forall i \in \mathcal{N} \label{9(c)}\\
& \qquad  c_i-\lambda_i-\underline{\tau_i}+\overline{\tau_i}=0,\forall i \in \mathcal{N} \label{9(d)}\\
& \qquad \overline{\mu_i}\geq0,\underline{\mu_i}\geq0,\overline{\tau_i}\geq0,\underline{\tau_i}\geq0,\overline{\eta_{ij}}\geq0,\underline{\eta_{ij}}\geq0.\label{9(e)}
\end{align}
\end{subequations}

Given $\underline{\tau_i}\geq0$, \eqref{9(d)} can be further simplified as $c_i-\lambda_i+\overline{\tau_i} \geq 0$. With the dual problem of IDSO in \eqref{9}, we model the interaction between IDSO and ITSO by a bilevel program in the next Section.

\section{Optimal Electricity Pricing Strategy}
In Section \ref{Problem Formulation}, we formulate the interaction between IDSO and ITSO as a bilevel program and point out the key difficulties of solving it. As one critical step of the proposed solution strategy, we derive the charging demand function in Section \ref{The Derivation of Price-Charging Demand Policy}. With it, we propose an effective solution strategy for the bilevel program in Section \ref{Solution Strategy}.

\subsection{Formulation of the Bilevel Power-Traffic Optimization}
\label{Problem Formulation}
The studied interaction problem between distribution and traffic networks has a bilevel structure, where the problem of ITSO in \eqref{7} is nested in the problem of IDSO in \eqref{9}. The bilevel program is formulated as follows
\begin{equation}\label{10}
\begin{split}
&\mathop{\min}_{\bm{\overline{\tau}},\bm{\underline{\tau}},\bm{\overline{\mu}},\bm{\underline{\mu}},\bm{\lambda},\bm{\underline{\eta}},\bm{\overline{\eta}}} \sum_{i \in \mathcal{N}}\overline{\mu_i}\cdot\overline{v_i}-\underline{\mu_i}\cdot\underline{v_i}+\overline{\tau_i}\cdot\overline{g_i}+  \\
& \qquad \ \
\sum_{i\in\mathcal{N}}\sum_{j \in \Omega_i}(f_{ij}\overline{\eta_{ij}}+f_{ij}\underline{\eta_{ij}})-\sum_{i \in \mathcal{N}}\lambda_i\cdot l_i-\sum_{i \in \mathcal{C}}\lambda_i\cdot d_i^* \\
& s.t. \eqref{9(b)},\eqref{9(c)},\eqref{9(d)},\eqref{9(e)}\\
& \quad \{d_i^*\}_{i \in \mathcal{C}} \in \mathop{\arg\min}_{\bm{\xi},\bm{f}}\sum_{a \in \mathcal{A}^e} \gamma \cdot \xi_a \cdot \tau_a(\xi_a) + \sum_{i \in \mathcal{C}} \lambda_i \cdot d_i\\
& \qquad \qquad \quad \ s.t. \eqref{6}, \eqref{7(c)},\eqref{7(d)},\eqref{7(e)}
\end{split}
\end{equation}
where the maximization problem of IDSO in \eqref{9} is replaced with an equivalent minimization problem by minimizing over the negative objective in \eqref{9(a)}. Once the bilevel program in \eqref{10} is solved, the electricity price for EV charging is obtained from the optimal solution of the dual variable $\lambda_i^*, \forall i \in \mathcal{C}$. 

Solving the bilevel optimization problem in \eqref{10} is typically done by replacing the lower-level problem with its KKT conditions for a single-level reduction \cite{BilevelReview}. However, this is not a trivial task. The presence of a bilinear term $\sum_{i \in \mathcal{C}}\lambda_i\cdot d_i^*$ in the objective function introduces nonlinearity, making the problem computationally challenging to solve with the optimality guarantee. Additionally, the IDSO and ITSO are separate entities, and the IDSO may not possess a detailed model of the ITSO \cite{Optpricing}, which can make it difficult to derive the KKT conditions necessary for solving \eqref{10} in practice.

Fortunately, the dual variable $\lambda_i, \forall i \in \mathcal{C}$ in the upper-level problem serves as the parameter in the lower-level problem of ITSO, where the charging demand $d_i,\forall i \in \mathcal{C}$ is the primal variable. The relationship between the price $\lambda_i, \forall i \in \mathcal{C}$ and the optimal charging demand $d_i^{*},\forall i \in \mathcal{C}$ is bijective and piecewise linear \cite{LinearComp}. We will explicitly derive this function in Section \ref{The Derivation of Price-Charging Demand Policy}. For now,
let's denote such bijection as the charging demand function, 
\begin{equation}\label{11}
\bm{d}^*_c=\pi(\bm{\lambda}_c), 
\end{equation}
where $\bm{d}^*_c=[d^*_i]_{i \in \mathcal{C}},\bm{\lambda}_c=[\lambda_i]_{i \in \mathcal{C}}$ are the vectors of charging demands and prices at the charging stations, respectively. Once such a function is obtained, the bilinear term $\sum_{i \in \mathcal{C}} \lambda_i \cdot d_i^*$ in the objective can be transformed to a term regarding the charging prices $\bm{\lambda}_c$, i.e., $(\bm{\lambda}_c)^\top \pi(\bm{\lambda}_c)$. 
 As such, IDSO can solve its own problem in \eqref{9} to determine the electricity price, 
without the need of knowing the detailed decision-making problem of ITSO. Therefore, deriving the function of $\pi(\bm{\lambda}_c)$ in \eqref{11} is the key step. The detailed solution process will be discussed in the next subsection.

\subsection{Derivation of Charging Demand Function}\label{The Derivation of Price-Charging Demand Policy}
The goal of this subsection is to derive the charging demand function between the vector of electricity price $\bm{\lambda}_c$ and the optimal charging demand $\bm{d}_c^*$ \textcolor{black}{by multiparametric programming theory \cite{LinearComp,Gal+2010}.} 
Before going into the details, we first rewrite the decision problem in \eqref{7} into a compact form of QP. Let $\bm{m}\in\mathbb{R}^{|\mathcal{W}|}=[m_w]_{m \in \mathcal{W}}$ be the vector formed by the demands of O-D pairs, and $\bm{E}\in \mathbb{R}^{|\mathcal{W}|\times\sum_{w \in \mathcal{W}}|\mathcal{R}_w|}$ be the coefficient matrix, we rewrite \eqref{7(c)} by $\bm{E}\bm{f}=\bm{m}$. Let $\bm{A} \in \mathbb{R}^{|\bm{A}^e|\times\sum_{w \in \mathcal{W}}|\mathcal{R}_w|}=[\bm{A}_1,...,\bm{A}_{|\mathcal{W}|}]$ denote the matrix formed by the horizontal stack of the link matrix of each O-D pair in the set $\mathcal{W}$. Therefore, \eqref{7(d)} can be rewritten as $\bm{A}\bm{f}=\bm{\xi}$. Denote $\bm{G}\in\mathbb{R}^{2|\mathcal{A}^e|\times|\mathcal{A}^e|}$ as the matrix formed by the vertical stack of the negative and positive values of the identity matrix $\bm{I}_{|\mathcal{A}^e|}$. $\bm{h}\in\mathbb{R}^{2|\mathcal{A}^e|}$ is the vertical stack of an all-zero vector and $\overline{\bm{\xi}}$. The constraint \eqref{7(e)} can be rewritten as $\bm{G}\bm{\xi}\leq\bm{h}$. To sum up, the compact QP form of \eqref{7} follows,
\begin{subequations}\label{12}
\begin{align}
& \mathop{\min}_{\bm{\xi},\bm{f}}\frac{1}{2}\bm{\xi}^\top\bm{Q}\bm{\xi}+\bm{q}^\top\bm{\xi}\label{12(a)}\\
& s.t. \ \bm{E}\bm{f}=\bm{m}:\bm{\psi}\label{12(b)}\\\
& \quad \ \
\bm{G}\bm{\xi} \leq \bm{h}:\bm{\phi}\label{12(c)}\\
& \quad \ \
\bm{A}\bm{f}=\bm{\xi}:\bm{\delta}\label{12(d)},
\end{align}  
\end{subequations}
where $\bm{q}$ is a column vector, whose first $|\mathcal{A}|$ elements equal $\gamma \cdot \xi_a^0,\forall a \in \mathcal{A}$, and the last $|\mathcal{C}|$ elements equal $\gamma(\xi_a^0+e_a/\rho_a)+e_a\cdot\lambda_i,\forall a \in \cup_{i\in\mathcal{C}}l_i^c$, while the remaining $|\mathcal{C}|$ elements equal zero. The detailed form of matrices $\bm{Q}, \bm{E}, \bm{G}, \bm{h}$ are given in Appendix \ref{Appendix B} for completeness. The charging prices $\bm{\lambda}_c$ are in the last $|\mathcal{C}|$ elements of the coefficient $\bm{q}$ which lies in the linear part of the objective. The charging demand is related to the flow $\xi_a$ on the virtual arc $l_i^c,\forall i \in \mathcal{C}$. We now proceed with the following proposition to characterize the \emph{local} relationship between the price $\bm{\lambda}_c$ and the optimal solutions.

\begin{proposition}\label{prop1}
Consider the convex QP problem \eqref{12}. Let $\bm{\xi}_i,\bm{f}_i,\bm{\psi}_i,\bm{\phi}_i,\bm{\delta}_i$ denote the local mapping between the charging price $\bm{\lambda}_c$ and the optimal solutions of primal and dual variables, $\Tilde{\bm{\xi}},\Tilde{\bm{f}},\Tilde{\bm{\psi}},\Tilde{\bm{\phi}},\Tilde{\bm{\delta}}$ denote the optimal primal and dual solutions under the given point $\Tilde{\bm{\lambda}}_c^i$. The calculation of $\bm{\xi}_i,\bm{f}_i,\bm{\psi}_i,\bm{\phi}_i,\bm{\delta}_i$, in the neighborhood of $\Tilde{\bm{\lambda}}_c^i$, is given by,
\begin{equation}\label{13}
\begin{bmatrix}
\bm{\xi}_i\\
\bm{f}_i\\
\bm{\psi}_i\\
\bm{\phi}_i\\
\bm{\delta}_i
\end{bmatrix}=
\begin{bmatrix}
\Tilde{\bm{\xi}}\\
\Tilde{\bm{f}}\\
\Tilde{\bm{\psi}}\\
\Tilde{\bm{\phi}}\\
\Tilde{\bm{\delta}}
\end{bmatrix}
+(\bm{M}_0)^{-1}\bm{N}_0(\bm{\lambda}_c-\Tilde{\bm{\lambda}}_c^i),
\end{equation}
where $\bm{N}_0 \in \mathbb{R}^{(4|\mathcal{A}^e|+\sum_{w \in \mathcal{W}}|\mathcal{R}_w|+|\mathcal{W}|)\times |\mathcal{C}|}$ and $\bm{M}_0 \in \mathbb{R}^{(4|\mathcal{A}^e|+\sum_{w \in \mathcal{W}}|\mathcal{R}_w|+|\mathcal{W}|)\times(4|\mathcal{A}^e|+\sum_{w \in \mathcal{W}}|\mathcal{R}_w|+|\mathcal{W}|)}$ that are defined as follows, 
\begin{align*}
&\bm{N}_0=\begin{bmatrix}
\bm{O} &
-\bm{J} & 
\bm{O} & 
\bm{O} & 
\bm{O} & 
\bm{O} 
\end{bmatrix}^\top;\\
&\bm{M}_0= \begin{bmatrix}
\bm{Q} & \bm{O} & \bm{O} & \bm{G}^\top &  -\bm{I}_{|\mathcal{A}^e|}\\
\bm{O} & \bm{O} & \bm{E}^\top & \bm{O} &  \bm{A}^\top\\
\bm{O} & \bm{E} & \bm{O} & \bm{O} & \bm{O} \\
-\bm{I}_{|\mathcal{A}^e|} & \bm{A} & \bm{O} & \bm{O} & \bm{O}\\
D(\Tilde{\bm{\phi}})\bm{G} & \bm{O} & \bm{O} & D(\bm{G}\Tilde{\bm{\xi}}-\bm{h}) & \bm{O}
\end{bmatrix}
\end{align*}
where $\bm{O}$ is an all-zero matrix, and $\bm{J}\in \mathbb{R}^{|\mathcal{C}|\times|\mathcal{C}|}$ is a diagonal matrix, whose diagonal element equals the average charging demand  $e_a$, and 
the operation $D(\cdot)$ creates a diagonal matrix from a vector. 
\end{proposition}

The proof of Proposition~\ref{prop1} is given in Appendix \ref{Appendix C}. Once \eqref{13} is obtained, we can use the mapping of $\bm{\xi}_i$ to obtain  $\pi_i(\bm{\lambda}_c)$, which is the local relationship between the optimal charging demand $\bm{d}_c^*$ and price $\bm{\lambda}_c$, i.e., 
\begin{equation}\label{14}
 \bm{d}_c^* = [\bm{O}\in \mathbb{R}^{|C|\times(|A|+|C|)},\bm{J}] \bm{\xi}_i,
\end{equation}
where $\bm{O}$ is an all-zero matrix with the size of $|C|\times(|A|+|C|)$.

For a given price point $\Tilde{\bm{\lambda}}_c^i$, in the neighborhood of $\Tilde{\bm{\lambda}}_c^i$, where \eqref{13} and \eqref{14} hold, a critical region $\mathcal{B}_i(\Tilde{\bm{\lambda}}_c^i)$ is defined that is related to a specific combination of the active constraints. 
Let $\hat{\bm{G}},\hat{\bm{h}}$ denote the coefficients of inactive constraints in \eqref{12(c)}, and $\hat{\bm{\phi}}_i$ be part of the $\bm{\phi}_i$ corresponding the active inequalities. By substituting $\bm{\xi}_i$ into the inactive constraints, feasibility is 
ensured. The optimality is ensured by $\hat{\bm{\phi}}_i \geq0$. The critical region of $\mathcal{B}_i(\Tilde{\bm{\lambda}}_c^i)$ is defined as
\begin{equation}\label{15}
\begin{split}
&\Tilde{\mathcal{B}}_i(\Tilde{\bm{\lambda}}_c^i)=\{\bm{E}\bm{f}_i=\bm{m}, \hat{\bm{G}}\bm{\xi}_i \leq \hat{\bm{h}},\bm{A}\bm{f}_i=\bm{\xi}_i,\hat{\bm{\phi}}_i \geq\bm{0},
\Lambda \}\\
&\mathcal{B}_i(\Tilde{\bm{\lambda}}_c^i)=\triangle \{\Tilde{\mathcal{B}}_i(\Tilde{\bm{\lambda}}_c^i)\},
\end{split}
\end{equation}
where $\Lambda$ denotes the given initial convex set of all feasible vectors of charging prices, and $\triangle \{\cdot\}$ is defined as an operator which removes redundant constraints. 

Let $\{\mathcal{B}_1(\Tilde{\bm{\lambda}}_c^1),...,\mathcal{B}_N(\Tilde{\bm{\lambda}}_c^N)\}$  be the set of all critical regions, then
\begin{equation}\label{16}
\begin{aligned}
&\mathcal{B}_i(\Tilde{\bm{\lambda}}_c^i) \cap  \mathcal{B}_j(\Tilde{\bm{\lambda}}_c^j)=\emptyset,\forall i \neq j\\
&\cup_{i=1}^N\mathcal{B}_i(\Tilde{\bm{\lambda}}_c^i)=\Lambda
\end{aligned}
\end{equation}

\begin{figure}
  \centering
  \includegraphics[scale=0.45]{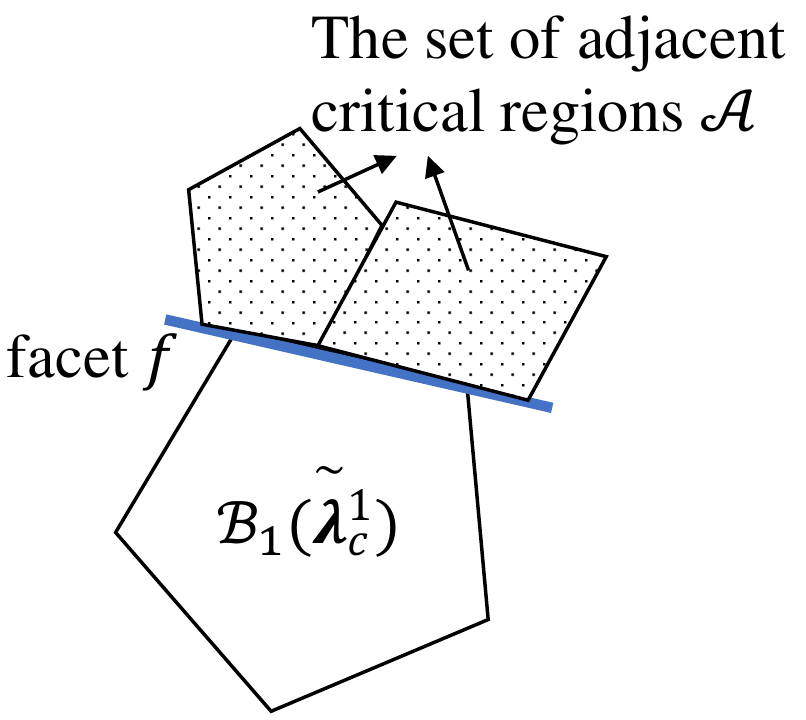}\\
  \caption{ The illustration of the enumeration of the critical regions, where $\mathcal{B}_1(\Tilde{\bm{\lambda}}_c^1)$ is an initial critical region whose one of the facets is $f$. The adjacent critical regions along this facet form the set $\mathcal{A}$.}\label{Fig 3}
\end{figure}

The goal is to compute all the critical regions via \eqref{15}, and the corresponding local policy by \eqref{14}. The exploration begins from each facet of the initial critical region, and all adjacent critical regions intersecting the facet are enumerated. \textcolor{black}{An illustration of such process is shown in Fig. \ref{Fig 3}.} In turn, each of these adjacent critical regions is considered and so on, until there are no more regions to be discovered. The main steps are summarized in Algorithm \ref{alg1}. 

\begin{algorithm}[h]
	\caption{The derivation of charging demand function $\pi(\bm{\lambda}_c)$ via multiparametric programming}
	\label{alg1}
	\begin{algorithmic}[1]
    \Require{The initial set of all feasible vectors of charging price $\Lambda$, the initial given point $\Tilde{\lambda}_c^1$, and the corresponding critical region $\mathcal{B}_1(\Tilde{\bm{\lambda}}_c^1)$;} 
    \State{Calculate the local function by \eqref{13},\eqref{14} and initialize the set of unexplored region $\mathcal{L}_{un}$ and the set of discovered region $\mathcal{L}_{ds}$ by $\mathcal{L}_{un} \longleftarrow \{\mathcal{B}_1(\Tilde{\bm{\lambda}}_c^1)\}$, $\mathcal{L}_{ds} \longleftarrow \{\mathcal{B}_1(\Tilde{\bm{\lambda}}_c^1)\}$}.
    \While{ $\mathcal{L}_{un}$ is not empty}.
    \State{Select and remove any region $\mathcal{R}$ in the set $\mathcal{L}_{un}$} 
    \For{Each facet $f$ of $\mathcal{R}$}
    \State{Compute set of critical regions $\mathcal{A}$ that along facet $f$, $\mathcal{L}_{un} \longleftarrow \mathcal{L}_{un}\cup(\mathcal{A}\backslash\mathcal{L}_{ds})$, $\mathcal{L}_{ds} \longleftarrow \mathcal{L}_{ds}\cup \mathcal{A}$}

    \State{Calculate the local function by \eqref{13},\eqref{14} for each critical region in the set $\mathcal{A}$}
    \EndFor
    \EndWhile
    \State{Collect all the solutions of local function and unify the critical regions $\mathcal{L}_{ds}$}
	\end{algorithmic}

\end{algorithm}

In a nutshell, the charging demand function $\pi(\bm{\lambda}_c)$ on the region of $\Lambda$ is a piecewise linear function, where the local function $\pi_i(\bm{\lambda}_c)$ is affine on the corresponding critical region $\mathcal{B}_i(\Tilde{\bm{\lambda}}_c^i)$. With this representation of $\bm{d}_c^* = \pi(\bm{\lambda}_c)$, the price setting problem of IDSO can be solved separately on each critical region with the local policy. The solution process is detailed in the next subsection.

\subsection{Solution Strategy}\label{Solution Strategy}
With the charging demand function $\pi(\bm{\lambda}_c)$, and the set of critical regions, we \textcolor{black}{use the half-space representation of a polytope and rewrite each critical region $\mathcal{B}_i(\Tilde{\bm{\lambda}}_c^i)$ as  $\bm{R}_i\cdot\bm{\lambda}_c\leq \bm{r}_i$}, whose  corresponding local function is $\pi_i(\bm{\lambda}_c)$. 
Thus, the problem of IDSO becomes a QP problem in each local region, which is given by 
\begin{subequations}\label{17}
\begin{align}
&\mathop{\min}_{\bm{\overline{\tau}},\bm{\underline{\tau}},\bm{\overline{\mu}},\bm{\underline{\mu}},\bm{\lambda},\bm{\underline{\eta}},\bm{\overline{\eta}}} \sum_{i \in \mathcal{N}}\overline{\mu_i}\cdot\overline{v_i}-\underline{\mu_i}\cdot\underline{v_i}+\overline{\tau_i}\cdot\overline{g_i}+  \label{17a}\\
& \qquad
\sum_{i\in\mathcal{N}}\sum_{j \in \Omega_i}(f_{ij}\overline{\eta_{ij}}-f_{ij}\underline{\eta_{ij}})-\sum_{i \in \mathcal{N}}\lambda_i\cdot l_i-(\bm{\lambda}_c)^\top \cdot \pi_i(\bm{\lambda}_c) \nonumber \\
& s.t. \eqref{9(b)},\eqref{9(c)},\eqref{9(d)},\eqref{9(e)}\nonumber\\
& \quad \  \bm{R}_i\cdot\bm{\lambda}_c\leq \bm{r}_i
\end{align}
\end{subequations}

We solve the QP problem in \eqref{17} for every local critical region 
and return the optimal solution of the problem, whose optimal objective is the smallest among all the problems. The corresponding optimal solution of $\bm{\lambda}^*$ represents the optimal electricity price across all power network nodes, part of which is the optimal charging price, i.e., $\bm{\lambda}_c^*=[\lambda_i^*]_{i \in \mathcal{C}}$.

\textbf{Remark 2:} Here, (17) is solved at single-period. For example, it can be solved every 15 minutes, or one hour. We note that the proposed approach can be extended to a multi-period problem, which minimizes the charging expenses, i.e., the last term of 17(a), over time. For that, the charging demand at each period is also replaced with the charging demand function at that period, which is affine on its critical region. In this line, the total charging expense over time is a quadratic term regarding the charging prices as well. The resulting problem is still a QP and can be solved by many off-the-shelf solvers.

Also, the OPF problem at the upper-level is replaced with its dual problem in \eqref{17}, which requires the OPF problem to be a convex one to make the strong duality hold. In this line, more realistic distribution network models can be considered, such as the one considering multiphase couplings and unbalanced loads, once there is a convex relaxation for it.

\subsection{Relationship with Other Modeling Perspectives}
\label{sec:equivalence}
Here, we discuss the relationship between this approach and others modeling the interaction between IDSO and ITSO from collaborative \cite{Optpricing,8737720} and game-theoretical \cite{ZHOU2021116703} perspectives.

\textit{1) Collaborative Perspective:} Ref. \cite{Optpricing,8737720} proposed to model the power and traffic network interaction by an joint optimization approach, i.e.,

\begin{equation}\label{18}
\begin{aligned}
&\mathop{\min}\ \eqref{8(a)}+\eqref{7(a)}-\sum_{i \in \mathcal{C}}\lambda_i\cdot d_i\\
&s.t. 
 \text{IDSO constraints \eqref{8(b)}-\eqref{8(e)}}\\
&\quad \ \text{ITSO constraints \eqref{7(b)}-\eqref{7(e)}}
\end{aligned}
\end{equation}

\textit{2) Game-theoretical Perspective:} Assuming generators and EVs are price-takers (as shown in Fig. \ref{Fig 1}), \cite{ZHOU2021116703} proposed that the interaction between the power and traffic networks can be modeled as a game with generators and EVs acting as the players to maximize their own profits. The equilibrium in the coupled
power and traffic networks can be described by combining the KKT conditions of the two problems, i.e.,

\begin{subequations}\label{19}
\begin{align}
&\text{KKT conditions of IDSO problem \eqref{8}}\label{19a}\\
&\text{KKT conditions of ITSO problem \eqref{7}}\label{19b}
\end{align}
\end{subequations}

\begin{figure}
  \centering
  \includegraphics[scale=0.45]{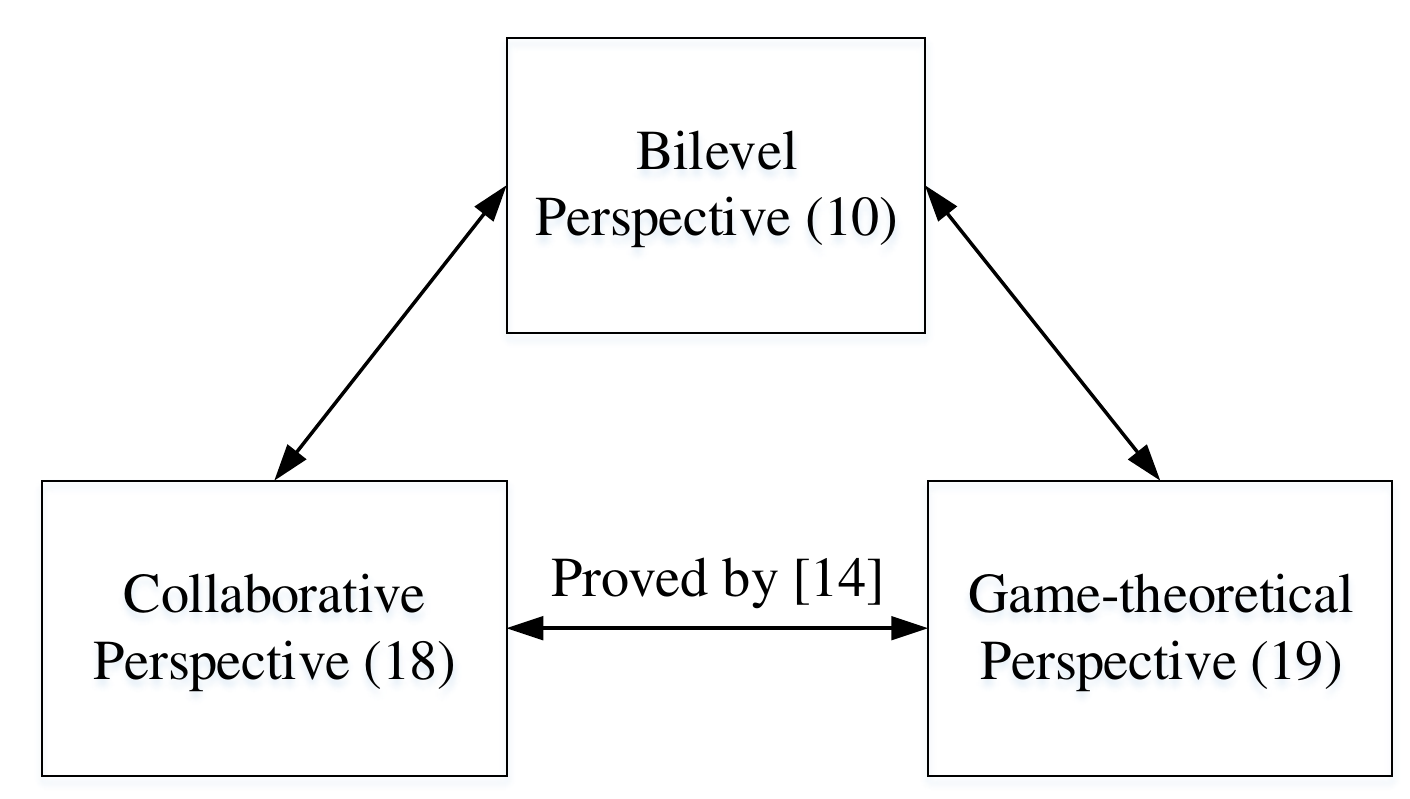}\\
  \caption{ The relationship between bilevel, collaborative and game-theoretical perspectives.}\label{Fig 4}
\end{figure}

By proving the KKT conditions of \eqref{18} have the same form as that of \eqref{19}, \cite{ZHOU2021116703} showed that the game-theoretical and collaborative perspectives are equivalent. In this work, we will show that our bilevel program has the same KKT conditions as that of \eqref{19}, and therefore is equivalent to the model from the game-theoretical perspective. The proof can be found in Appendix \ref{Appendix D}. Leveraging the relationship established in \cite{ZHOU2021116703} that the game-theoretical and collaborative perspectives are equivalent, our model is also equivalent to the collaborative perspective. The relationship is summarized in Fig. \ref{Fig 4}.


\section{Case Study}
\begin{figure}
  \centering
  \includegraphics[scale=0.4]{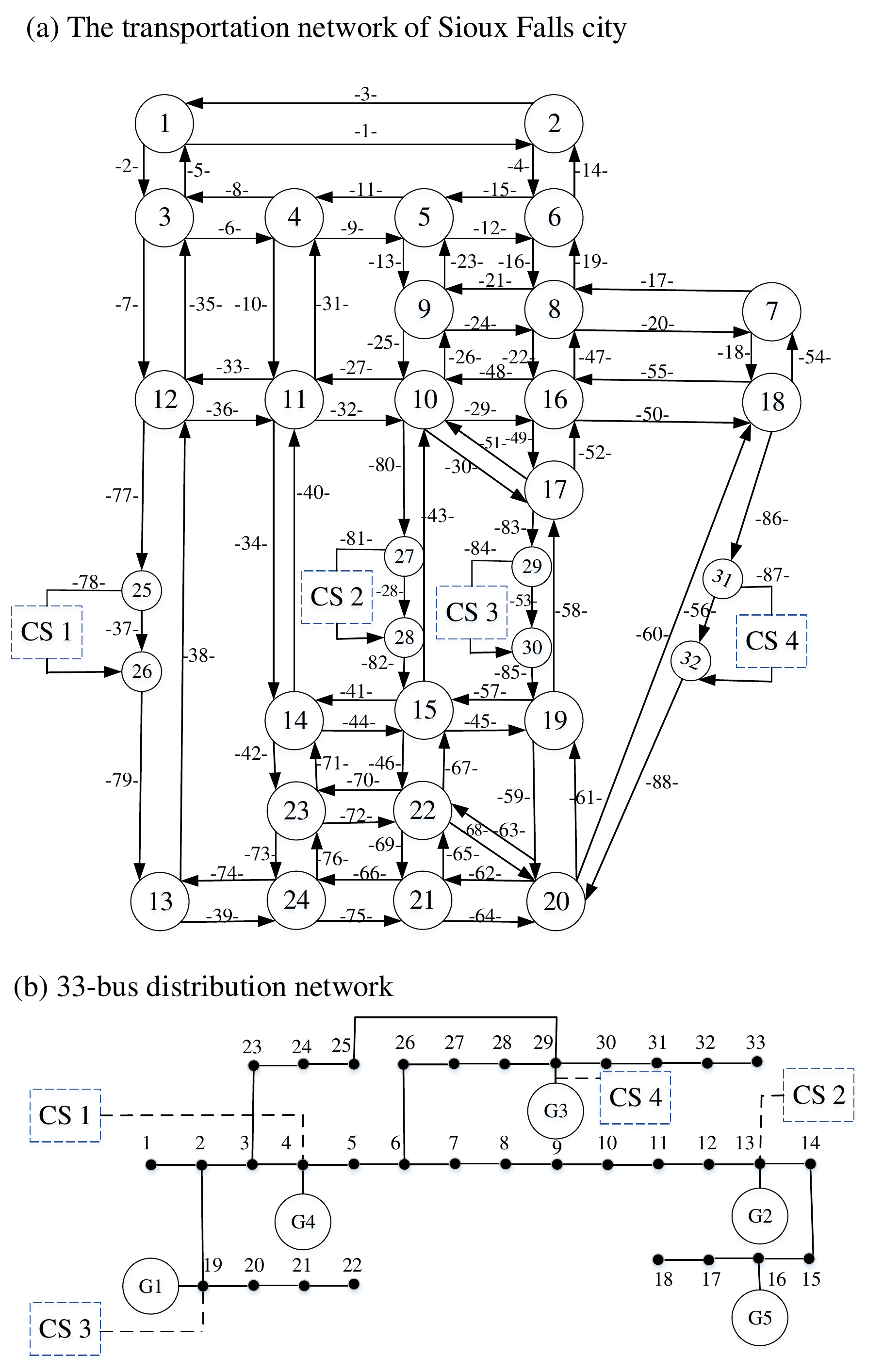}\\
  \caption{ The Sioux Falls city traffic network and 33-bus distribution network coupled by charging stations (CSs).}\label{Fig 5}
\end{figure}

This section presents the results of testing the effectiveness of the proposed approach on a test system~\cite{9163146} consisting of the traffic network of Sioux Falls city and the 33-bus distribution network~\cite{25627}, as shown in Fig. \ref{Fig 5}.
The coupled networks include four charging stations with specified flow capacity $\overline{\xi_a}$ and charging demand capacity, as outlined in Table \uppercase\expandafter{\romannumeral1}. In the traffic network, the O-D pairs of Nodes \#1-\#13 and Nodes \#1-\#20 are considered for EVs and Nodes \#2-\#13 and Nodes \#2-\#20 are considered for regular vehicles (RVs). Although our approach is applicable to the consideration of more O-D pairs, we assume that there are only four pairs having the traffic demand, while others don't. The summary of routes in each O-D pair can be found in \cite{USDR2020}. The shape parameters in \eqref{5} and the cost of unit time are given as $\xi_a^0=0, R_a=10^4,\gamma=10^3$. The average charging demand $e_a$ of an individual vehicle is set as 12 kWh, and the charging rate $\rho_a$ is 200 kW. 
In the distribution network, the lower and upper bounds of voltage are set as 0.9 p.u. and 1.1 p.u., respectively. There are 5 distributed generators (G1 - G5) connected to the buses \#4,\#13,\#16,\#19,\#29, as shown in Fig \ref{Fig 5}. The capacity and the cost coefficient of each generator are given in Table \uppercase\expandafter{\romannumeral2}. Also, the fixed loads of IEEE 33 network are scaled up to ten times.

\begin{table}[ht]
\caption{ The flow capacity and charging demand capacity for the four EV charging stations.}
\begin{center}
\begin{tabular}{c c c c  c }
\hline\hline  &\makecell[c]{Charging\\ station 1} & \makecell[c]{Charging\\ station 2} &
\makecell[c]{Charging\\ station 3} & \makecell[c]{Charging\\ station 4}\\
\hline
\makecell[c]{Flow capacity \\ (number of vehicles)} &100 & 250 & 250 & 100\\
\hline
\makecell[c]{Charging demand \\ capacity (kWh)}&1200 & 3000 & 3000 & 1200\\
\hline\hline
\end{tabular}
\end{center}
\end{table}
\begin{table}[ht]
\caption{Generation capacity and cost for the five generators.}
\begin{center}
\begin{tabular}{c c c c  c c}
\hline\hline  &\makecell[c]{G1} & \makecell[c]{G2} &
\makecell[c]{G3} & \makecell[c]{G4}& \makecell[c]{G5}\\
\hline
\makecell[c]{Generation capacity (kWh)}&$10^4$ & $10^4$ & $3 \times 10^4$ & $10^4$& $10^4$\\
\hline
\makecell[c]{Cost coefficient (\$/kWh)}&0.8 & 0.6 & 0.5 & 0.7 &0.4\\
\hline\hline
\end{tabular}
\end{center}
\end{table}

We conduct the experiments from three angles: (1) To investigate the impact of O-D pair traffic demand on the charging demand; 
(2) To show the effectiveness of the charging price on affecting the charging load patterns; 
(3) To show the bilevel model is robust against the traffic demand forecast error to some extent. 
All simulations are implemented on the laptop with Intel®CoreTM i5-10210U 1.6 GHz CPU, and 8.00 GM RAM and based on the Multiparametric Toolbox \cite{MPT3}.

\subsection{Impacts of Traffic Demands on Charging Demands and Prices}
We first assign each O-D pair with equal traffic demand and test the proposed method under different levels of it, namely 100, 200, and 300 vehicles. The cost of IDSO, the number of critical regions, and the corresponding computation time are shown in Table \uppercase\expandafter{\romannumeral3}. Unsurprisingly, the increase in traffic demand results in an increase of charging demand, and therefore, the operational cost of IDSO increases correspondingly. In addition, we observe that as a parameter in the equality constraint \eqref{7(c)} of the ITSO decision problem, the value of the traffic demand has impact on the traffic pattern and results in different numbers of critical regions. \textcolor{black}{Concretely, the number of critical regions reduces as the traffic demand becomes larger. With the larger demand, more inequality constraints regarding the arc flows are binding, which results in a smaller combination of operation patterns, and therefore a smaller number of critical regions.} 
Since the proposed approach iterates over all critical regions, the computation time increases when the number of critical regions is larger. Furthermore, we also calculate the cost of IDSO by solving the joint optimization in \eqref{18}, whose results are listed in the last row of Table \uppercase\expandafter{\romannumeral3}, and the cost of IDSO by solving \eqref{10} using single-level reduction via KKT conditions, whose computation time is also reported. Unsurprisingly, the costs of the IDSO by solving \eqref{18} and our bilevel program in 
\eqref{10} are the same, as they are equivalent problems shown in Section \ref{sec:equivalence}. However, \eqref{18} requires the whole decision model of the ITSO, while our approach doesn't and still achieves optimality. For the comparison candidate using single-level reduction via KKT conditions, it is transformed to a mixed integer program. Therefore, its computation time is much longer than the propose approach, which shows that the proposed approach is more computationally efficient.



\begin{table}[htbp]
\caption{  Under different traffic demand levels, the cost of IDSO solved by the proposed approach for the bilevel program in \eqref{10}, the number of critical regions, the computation time of the proposed approach, the computation time of solving \eqref{10} by single-level reduction via KKT conditions, the cost of IDSO of solving \eqref{10} by single-level reduction via KKT conditions, and the cost of IDSO by solving the joint optimization in \eqref{18}.}
\begin{center}
\begin{tabular}{c c c c}
\hline\hline  &$m_w=100$ & $m_w=200$ &$m_w=300$\\
\hline
\makecell[c]{The cost of IDSO (\$) \\
by the proposed approach}&19967 & 21478 & 23068\\
\hline
\makecell[c]{The number of the\\ critical regions}&93 & 8 & 3\\
\hline
\makecell[c]{Computation time of the \\
proposed approach (s)
}&14 & 0.86 & 0.32\\
\hline
\makecell[c]{Computation time of \\solving \eqref{10} via \\
KKT conditions (s)
}&145 & 5640 & 44\\
\hline
\makecell[c]{The cost of IDSO (\$) of \\solving \eqref{10} via \\
KKT conditions}&19967 & 21478 & 23068\\
\hline
\makecell[c]{The cost of IDSO (\$) \\
by solving joint \\ optimization \cite{Optpricing,8737720}}&19967 & 21478 & 23068\\
\hline\hline
\end{tabular}
\end{center}
\end{table}
\begin{figure}[tbp]
  \centering
  \includegraphics[scale=0.65]{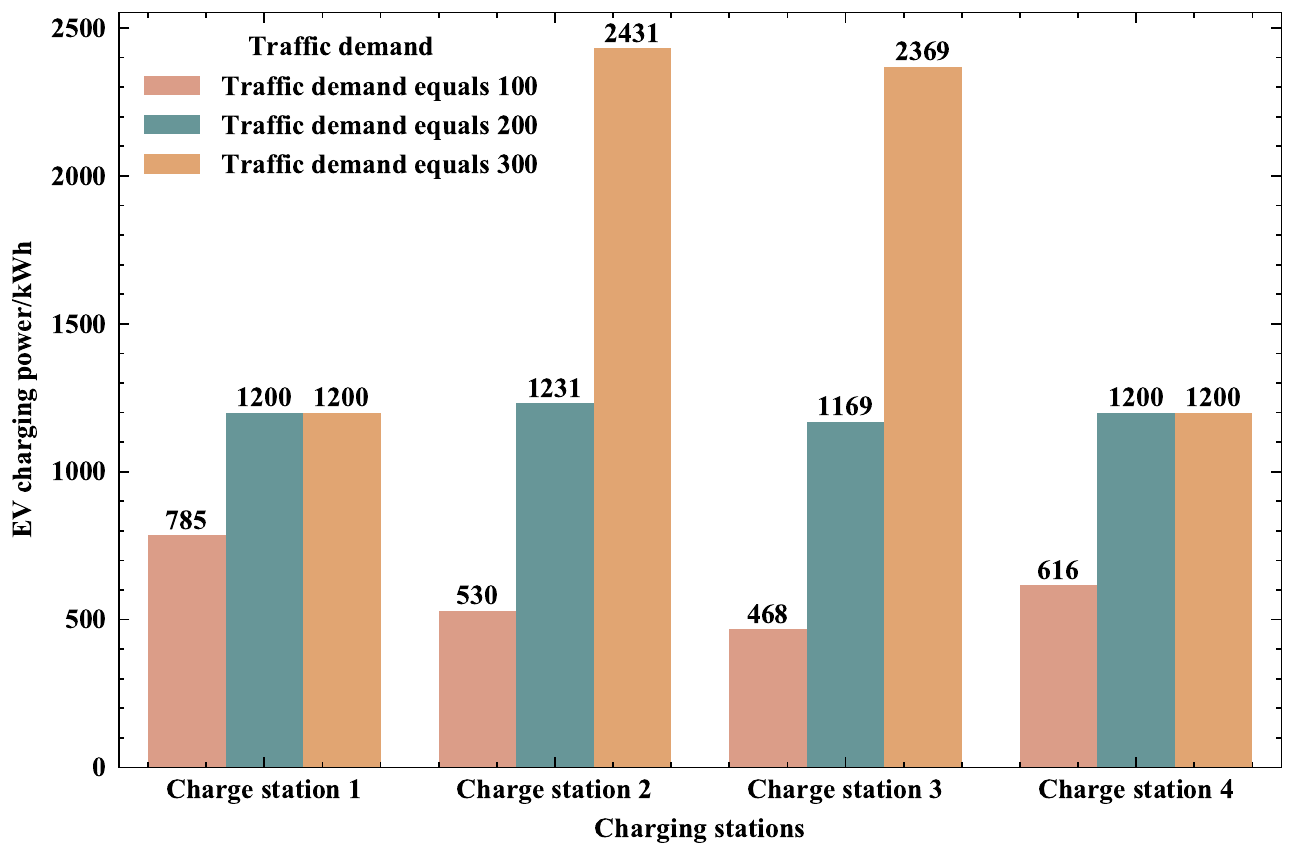}\\
  \caption{The charging demand at four charging stations under different traffic demand levels.}\label{Fig 6}
\end{figure}
We show the electricity price at each charging station and the charging demand under different traffic demand levels in the first row of Table \uppercase\expandafter{\romannumeral4} 
and Fig. \ref{Fig 6}, respectively. Since the charging price under different traffic demand levels remains the same, we show the results in a single line.
As the line connecting node \#29 and node \#25 is congested, the charging prices are different in different charging stations. It is shown in Fig. \ref{6} that the charging demands in the traffic network can be affected by the charging price. With a lower charging price, the charging demand in the charging station (CS) \#2 is always larger than the charging demand in CS \#3, across different traffic demand levels. Also, we observe that the charging price is not the only factor affecting the charging demands. The traffic network structure, such as the constituent routes of each O-D pair, has a significant impact as well. For instance, among the 6 routes of the EV O-D pair Node \#1-\#13, 4 routes pass CS \#1. Therefore, it carries most of the charging demand of O-D pair Node \#1-\#13. Hence, when the traffic demand of each O-D pair is 100 vehicles, the charging demand in CS \#1 is the largest. When the traffic demand of each O-D pair increases to 200 and 300 vehicles, the demand in CS \#1 remains operating at the maximum capacity.

\begin{table}[h]
\caption{The charging prices (\$/kWh) at four charging stations after the decrease in generation cost (\$/kWh) of G3.}
\begin{center}
\begin{tabular}{c c c c c}
\hline\hline  
\makecell[c]{Generation cost \\of G3 }&\makecell[c]{Charging \\station 1} &\makecell[c]{Charging \\station 2} & \makecell[c]{Charging \\station 3} & \makecell[c]{Charging \\station 4}\\
\hline
0.5&0.7&0.62 & 0.71 & 0.5\\
\hline
0.3&0.8&0.6 & 0.82 & 0.3\\
\hline
0.1&0.74&0.48 & 0.75 & 0.1\\
\hline
0&0.78&0.47 & 0.8 & 0\\
\hline\hline
\end{tabular}
\end{center}
\end{table}

In addition, we compare the proposed approach with a baseline charging strategy in which all EVs choose to charge at the charging stations with the lowest charging price, specifically at CS \#2 and CS \#4. We compare the costs of IDSO, ITSO, and the operation cost of both networks defined in the objective of \eqref{18}, which is the negative value of the social welfare \cite{Optpricing}. The results of this comparison when the traffic demand is set to 100 vehicles are presented in Table \uppercase\expandafter{\romannumeral5}.
\begin{table}[htbp]
\caption{The costs of IDSO, ITSO, and two networks' operation cost under the proposed approach and the baseline approach}
\begin{center}
\begin{tabular}{c c c c}
\hline\hline  & \makecell[c]{IDSO \\
cost (\$)} & \makecell[c]{ITSO \\
cost (\$)} &\makecell[c]{Two networks' \\operation cost (\$)}\\
\hline
\makecell[c]{The proposed approach}& 19967 & 43258 & 63225\\
\hline
\makecell[c]{Baseline: charging EVs at \\the lowest LMP nodes}& 19794 & 48294 & 68088\\
\hline\hline
\end{tabular}
\end{center}
\end{table}
Although the baseline charging strategy results in the lower operating cost of IDSO, the cost of ITSO is larger than the proposed approach, as the traffic time on arcs increases when all EVs travel to the same stations with lower prices for charging.
The two networks' operation cost of the baseline approach increases by 7.6\%, compared with the proposed bilevel optimization approach. 

Moreover, to show the proposed approach is applicable to large scale distribution network, we apply the proposed approach on the 85-bus \cite{das1995simple} and 136-bus \cite{mantovani2000reconfiguraccao} distribution networks, as well. The generators and the charging stations are connected to the nodes with the same indexes as the one shown in Fig. \ref{Fig 5}. The traffic demand of 100 vehicles is considered. The charging demand function is a piecewise linear function defined on 93 critical regions. We run the program for ten times under 33-bus, 85-bus and 136-bus distribution networks. The average computation time and its standard error are reported in Table \uppercase\expandafter{\romannumeral6}. The computation time increases as the scale of the network becomes larger, as a larger number of variables and constraints are considered. However, even for the case concerning the 136-bus network, the computation time is acceptable for our optimization problem, which is often conducted on a hourly basis.

\begin{table}[ht]
\caption{ The average and standard error of the computation time in ten runs under 33-bus, 85-bus and 136-bus distribution networks.}
\begin{center}
\begin{tabular}{c c c c}
\hline\hline  &33-bus network & 85-bus network &
136-bus network\\
\hline
Average  &14 s &63 s & 160 s \\
Standard error &2 s &1 s & 6 s \\
\hline\hline
\end{tabular}
\end{center}
\end{table}

\subsection{Impacts of Electricity Prices on Charging Demands}
To demonstrate the impact of charging prices on charging demand, we investigate the effect of generation costs on charging demand patterns. We can view the cost coefficient of the generator connected to a node as an equivalent cost coefficient of a generation mix. Therefore, with the increasing penetration of renewable sources, the cost of the generation mix decreases. For this scenario, we test different values of the cost coefficient of the generator G3 connected to node \#29, where CS \#4 is located, while keeping the other coefficients the same.  Specifically, we set the generation cost of G3 as 0.5, 0.3, 0.1, and 0 \$/kWh, respectively, and set the traffic demand of each O-D pair to 100 vehicles. The charging prices are outlined in Table \uppercase\expandafter{\romannumeral4}. As the cost coefficient of G3 decreases, the price difference between CS \#1 and CS \#4 becomes larger.

\begin{figure}
  \centering
  \includegraphics[scale=0.65]{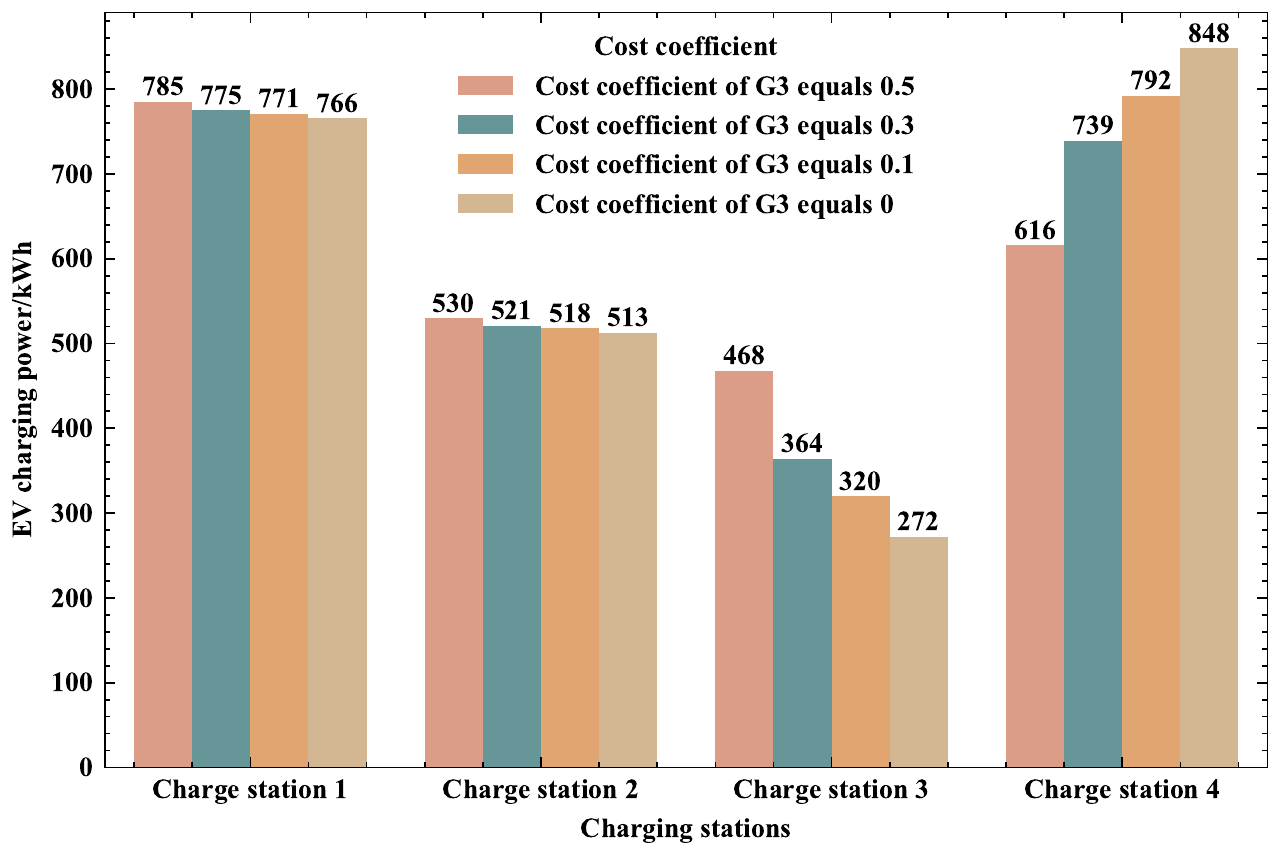}\\
  \caption{The charging demand at four charging stations under different generation costs of G3.}\label{Fig 7}
\end{figure}

Fig. \ref{Fig 7} illustrates the charging demands under different generation costs of G3. The impact of charging prices on charging demand is significant. When the cost coefficient of G3 is 0.5 \$/kWh and 0.3 \$/kWh, the charging demand at CS \#1 is higher than that at CS \#4. However, as the cost coefficient of G3 decreases (and thus the price at CS \#4 decreases), the charging demand at CS \#4 increases and eventually surpasses the charging demand at CS \#1. This shows that electricity prices can effectively alter the spatial distribution of charging demand and incentivize EVs to charge at locations with lower electricity prices, which may coincide with locations with plentiful renewable sources and the marginal generation cost is low.

\subsection{Investigation on Inaccurate Traffic Demand Forecasts}
\begin{figure}
  \centering
  \includegraphics[scale=0.65]{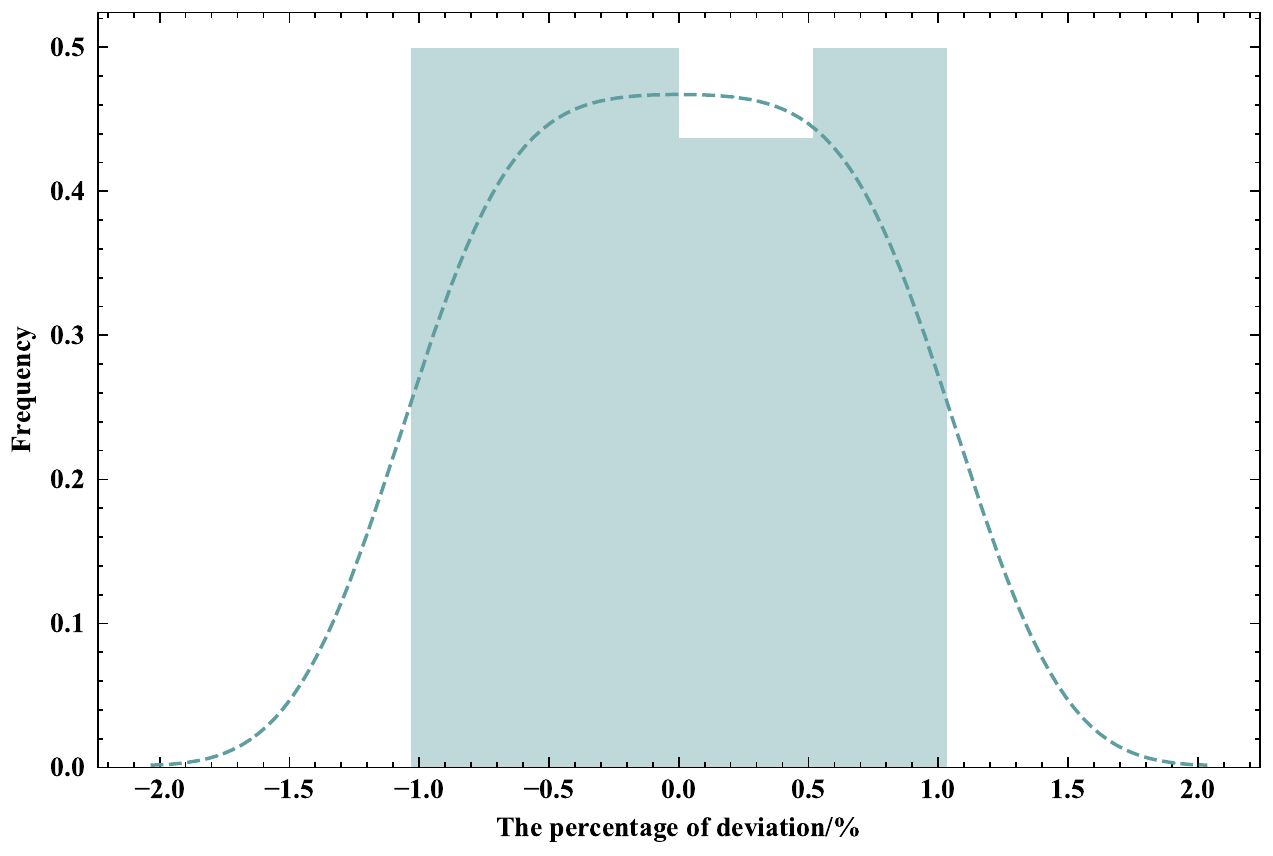}\\
  \caption{The distribution of the deviation between the IDSO's cost under the forecast and the realization.}\label{Fig 8}
\end{figure}

It is widely acknowledged that obtaining high-quality estimates or forecasts of traffic demand is challenging, and the actual traffic demand $m_w,\forall w \in \mathcal{W}$ may deviate from the forecast value. In this subsection, we test the robustness of the bilevel model against the forecast error. Specifically, we assume the accurate values of the traffic demand for each O-D pair are all equal to 300 vehicles. We generate 30
samples of the possible forecasts which are the integers ranging from 285 to 315 vehicles ($\pm 5\%$ deviations from the ground-truth traffic demand). 
The percentage of the deviation between the IDSO’s costs under the forecast and the realization is then calculated and shown in the histogram Fig. \ref{Fig 8}. The maximum cost deviation percentage is less than 1.5\%, which shows the IDSO’s costs under the forecast and the realization are close. Since the deviation of the traffic demand is in a relatively small range that doesn't change the charging patterns, the charging demand still reaches the maximum capacity of CS \#1 and CS \#4, and only the demands in the CS \#2 and CS \#3 are different under different forecast values. Also, as the charging price under different levels of charging demands remains the same, the different charging demands in the two stations have a small impact on the cost of IDSO. Therefore, the results show that the bilevel model is robust against the forecast error of traffic demand to some extent.

\section{Conclusion}
In this work, we propose to manage EV charging via a bilevel optimization approach, where the IDSO determines the charging price at the upper level, and the ITSO decides the optimal charging power for the given price at the lower level. The charging demand function depicting the theoretical relationship between the charging price and the optimal charging demand is derived, which is piecewise linear. With the charging demand function, the IDSO can solve the optimal EV management problem without the need for a detailed model of ITSO, while still considering the coupling of the power-traffic networks. In addition, with the bilinear term replaced with a piecewise quadratic term, we are able to solve the bilevel optimization efficiently with optimality guarantees. 
Case studies reveal that the different levels of traffic demand can affect the charging pattern in the charging stations. Also, charging prices can effectively change the charging demand, where the charging stations with lower price shoulder more charging demand than stations with higher prices. 

In the future, it would be interesting to extend our approach to multi-period EV charging optimization. Besides, the single-phase distribution network model is used here for simplicity. It is still required to explore more realistic distribution network model, considering the multiphase couplings and unbalanced loads. Additionally, our approach provides a novel solution to the bilevel program by leveraging the relationship between the upper-level variable and the optimal solution of the lower-level problem, allowing for the upper-level problem to be solved independently with an optimality guarantee. It is also interesting to apply the proposed approach to other tasks modeled as a bilevel program.

\section*{Acknowledgement}
The authors would like to appreciate the Shanghai Jiao Tong University Grants.

\bibliographystyle{IEEEtran}
\bibliography{IEEEabrv,mylib}

\appendices

\section{The Coefficients of the Decision-making of ITSO in Compact QP Form}\label{Appendix B}

$\bm{Q}$ is a diagonal matrix, whose the first $|\mathcal{A}|$ and the last $|\mathcal{C}|$ principal diagonal elements equal $2\gamma/R_a,\forall a \in \mathcal{A}\cup (\cup_{i\in\mathcal{C}}l_i^c)$, and the remaining ones equal zero. That is, the diagonal elements are $[\underbrace{\frac{2\gamma}{R_1},\hdots,\frac{2\gamma}{R_{|\mathcal{A}|}}}_{|\mathcal{A}|},\underbrace{0,\hdots,0}_{|\mathcal{C}|},\underbrace{\frac{2\gamma}{R_{l_1^c}},\hdots,\frac{2\gamma}{R_{l_{|\mathcal{C}|}^c}}}_{|\mathcal{C}|}]$. Also,
\begin{align*}
\bm{E} =
\begin{bmatrix}
\underbrace{1 \hdots 1}_{|\mathcal{R}_1|} \underbrace{0 \hdots 0 0 \hdots 0}_{\sum_{i=2}^{|\mathcal{W}|}|\mathcal{R}_i|}\\
\underbrace{0 \hdots 0}_{|\mathcal{R}_1|} \underbrace{1 \hdots 1}_{|\mathcal{R}_2|} \underbrace{0 \hdots 0}_{\sum_{i=3}^{|\mathcal{W}|}|\mathcal{R}_i|}\\
 \ddots \\
0 \hdots  \hdots  0 \underbrace{1 \hdots 1}_{|\mathcal{R}_{|\mathcal{W}|}|}
\end{bmatrix},
\bm{G} =
\begin{bmatrix}
-\bm{I}_{|\mathcal{A}^e|}\\
\bm{I}_{|\mathcal{A}^e|}
\end{bmatrix},
\bm{h} =
\begin{bmatrix}
\bm{0}_{|\mathcal{A}^e|}\\
\overline{\bm{\xi}}
\end{bmatrix}
\end{align*}

\section{Proof of Proposition 1}\label{Appendix C}

The Lagrangian of \eqref{12} is given by

\begin{equation}\label{22}
\begin{split}
& L=\frac{1}{2}\bm{\xi}^\top\bm{Q}\bm{\xi}+\bm{q}^\top\bm{\xi}+\bm{\psi}^\top(\bm{E}\bm{f}-\bm{m})+\bm{\phi}^\top(\bm{G}\bm{\xi}-\bm{h})+\\
&\qquad \qquad \quad \bm{\delta}^\top(\bm{A}\bm{f}-\bm{\xi})
\end{split}
\end{equation}

When $\bm{\lambda}_c=\Tilde{\bm{\lambda}}_c^i$, the optimal solution of primal and dual variables becomes $\Tilde{\bm{\xi}},\Tilde{\bm{f}},\Tilde{\bm{\phi}},\Tilde{\bm{\psi}},\Tilde{\bm{\delta}}$.  The KKT conditions for stationarity, primal feasibility, and complementary slackness are

\begin{subequations}\label{23}
\begin{align}
& \bm{Q}\Tilde{\bm{\xi}}+\bm{q}+\bm{G}^\top\Tilde{\bm{\phi}}-\Tilde{\bm{\delta}}=0\\
&\bm{E}^\top\Tilde{\bm{\psi}}+\bm{A}^\top\Tilde{\bm{\delta}}=0\\
&\bm{E}\Tilde{\bm{f}}-\bm{m}=0\\
&\bm{A}\Tilde{\bm{f}}-\Tilde{\bm{\xi}}=0\\
&D(\Tilde{\bm{\phi}})(\bm{G}\Tilde{\bm{\xi}}-\bm{h})=0,
\end{align}
\end{subequations}
where the operation $D(\cdot)$ creates a diagonal matrix from a vector.  Taking the differentials of these conditions gives the equations

\begin{subequations}\label{24}
\begin{align}
& d\bm{Q}\Tilde{\bm{\xi}}+\bm{Q}d\bm{\xi}+d\bm{q}+d\bm{G}^\top\Tilde{\bm{\phi}}+\bm{G}^\top d\bm{\phi}-\bm{I}_{|\mathcal{A}^e|}d\bm{\delta}=0\\
&d\bm{E}^\top\Tilde{\bm{\psi}}+\bm{E}^\top d\bm{\psi}+d\bm{A}^\top\Tilde{\bm{\delta}}+\bm{A}^\top d\bm{\delta}=0\\
&d\bm{E}\Tilde{\bm{f}}+\bm{E}d\bm{f}-d\bm{m}=0\\
&d\bm{A}\Tilde{\bm{f}}+\bm{A}d\bm{f}-\bm{I}_{|\mathcal{A}^e|}d\bm{\xi}=0\\
&D(\bm{G}\Tilde{\bm{\xi}}-\bm{h})d\bm{\phi}+D(\Tilde{\bm{\phi}})(d\bm{G}\Tilde{\bm{\xi}}+\bm{G}d\bm{\xi}-d\bm{h})=0
\end{align}
\end{subequations}

We rewrite \eqref{24} into a compact matrix form

\begin{equation}\label{25}
\begin{split}
&\begin{bmatrix}
\bm{Q} & \bm{O} & \bm{O} & \bm{G}^\top &  -\bm{I}_{|\mathcal{A}^e|}\\
\bm{O} & \bm{O} & \bm{E}^\top & \bm{O} &  \bm{A}^\top\\
\bm{O} & \bm{E} & \bm{O} & \bm{O} & \bm{O} \\
-\bm{I}_{|\mathcal{A}^e|} & \bm{A} & \bm{O} & \bm{O} & \bm{O}\\
D(\Tilde{\bm{\phi}})\bm{G} & \bm{O} & \bm{O} & D(\bm{G}\Tilde{\bm{\xi}}-\bm{h}) & \bm{O}
\end{bmatrix}\\
&\cdot
\begin{bmatrix}
d\bm{\xi}\\
d\bm{f}\\
d\bm{\psi}\\
d\bm{\phi}\\
d\bm{\delta}
\end{bmatrix}=
\begin{bmatrix}
-d\bm{Q}\Tilde{\bm{\xi}}-d\bm{q}-d\bm{G}^\top\Tilde{\bm{\phi}}\\
-d\bm{E}^\top\Tilde{\bm{\psi}}-d\bm{A}^\top\Tilde{\bm{\delta}}\\
-d\bm{E}\Tilde{\bm{f}}+d\bm{m}\\
-d\bm{A}\Tilde{\bm{f}}\\
D(\Tilde{\bm{\phi}})d\bm{h}-D(\Tilde{\bm{\phi}})d\bm{G}\Tilde{\bm{\xi}}
\end{bmatrix}
\end{split}
\end{equation}

The left-hand side coefficient matrix is the matrix $\bm{M}_0$. We wish to compute the Jacobian $\frac{\partial \bm{\xi}}{\partial \bm{\lambda}_c},\frac{\partial \bm{f}}{\partial \bm{\lambda}_c},\frac{\partial \bm{\psi}}{\partial \bm{\lambda}_c},\frac{\partial \bm{\phi}}{\partial \bm{\lambda}_c},\frac{\partial \bm{\delta}}{\partial \bm{\lambda}_c}$. And take $\frac{\partial \bm{\xi}}{\partial \bm{\lambda}_c}$ for instance, it can be calculated according to the chain rule, i.e.,$\frac{\partial \bm{\xi}}{\partial \bm{\lambda}_c}=\frac{\partial \bm{\xi}}{\partial \bm{q}}\cdot\frac{\partial \bm{q}}{\partial \bm{\lambda}_c}$. Therefore, we firstly calculate the Jacobian $\frac{\partial \bm{\xi}}{\partial \bm{q}},\frac{\partial \bm{f}}{\partial \bm{q}},\frac{\partial \bm{\psi}}{\partial \bm{q}},\frac{\partial \bm{\phi}}{\partial \bm{q}},\frac{\partial \bm{\delta}}{\partial \bm{q}}$ by substituting $d\bm{q}=\bm{I}_{|\mathcal{A}^e|}$, and setting all other differential terms in the right-hand side to zero. Therefore, the Jacobian is given by 

\begin{equation}\label{26}
\begin{split}
&\begin{bmatrix}
\frac{\partial \bm{\xi}}{\partial \bm{q}}&
\frac{\partial \bm{f}}{\partial \bm{q}}&
\frac{\partial \bm{\psi}}{\partial \bm{q}}&
\frac{\partial \bm{\phi}}{\partial \bm{q}}&
\frac{\partial \bm{\delta}}{\partial \bm{q}}
\end{bmatrix}^\top \\
&=\bm{M}_0^{-1}\cdot
\begin{bmatrix}
-\bm{I}_{|\mathcal{A}^e|}&
\bm{O}&
\bm{O}&
\bm{O}&
\bm{O}
\end{bmatrix}^\top
\end{split}
\end{equation}

The Jacobian of $\frac{\partial \bm{q}}{\partial \bm{\lambda}_c}$ equals

\begin{equation}\label{27}
\begin{bmatrix}
\bm{O}&
\bm{J} 
\end{bmatrix}^\top
\end{equation}
which is the stack of a zero matrix with the dimension of $(|\mathcal{A}|+|\mathcal{C}|)\times|\mathcal{C}|$ and a diagonal matrix $\bm{J}\in \mathbb{R}^{|\mathcal{C}|\times|\mathcal{C}|}$ whose principal diagonal element equals the average charging demand $e_a,\forall a \in \cup_{i \in \mathcal{C}}l_i^c$. By applying the chain rule, the Jacobian of $\frac{\partial \bm{\xi}}{\partial \bm{\lambda}_c},\frac{\partial \bm{f}}{\partial \bm{\lambda}_c},\frac{\partial \bm{\psi}}{\partial \bm{\lambda}_c},\frac{\partial \bm{\phi}}{\partial \bm{\lambda}_c},\frac{\partial \bm{\delta}}{\partial \bm{\lambda}_c}$ is

\begin{equation}\label{28}
\bm{M}_0^{-1}\cdot
\begin{bmatrix}
-\bm{I}_{|\mathcal{A}^e|}&
\bm{O}&
\bm{O}&
\bm{O}&
\bm{O}
\end{bmatrix}^\top
\cdot
\begin{bmatrix}
\bm{O}&
\bm{J}
\end{bmatrix}^\top
\end{equation}

We use $\bm{N}_0$ to denote the multiplication of the last two matrices, i.e.,

\begin{equation}\label{29}
\begin{bmatrix}
\bm{O}&
-\bm{J}&
\bm{O}&
\bm{O}&
\bm{O}&
\bm{O}
\end{bmatrix}^\top
\end{equation}

Therefore, the Jacobian in \eqref{28} is replaced with $\bm{M}_0^{-1}\bm{N}_0$.

\section{The Derivation of the KKT 
 Conditions of the Bilevel Program}\label{Appendix D}

Here, we derive the KKT conditions of the bilevel problem, where the IDSO decision problem is at the upper-level, and the ITSO decision problem is at the lower-level. We rewrite \eqref{10} by replacing the dual problem at the upper-level with the primal problem in \eqref{8}, i.e.,

\begin{equation}\label{30}
\begin{aligned}
&\mathop{\min}_{\bm{g},\bm{v},\bm{\theta}}\bm{c}^\top\bm{g}\\
& s.t. \eqref{8(b)},\eqref{8(c)},\eqref{8(d)},\eqref{8(e)}\\
& \quad \ \mathop{\min}_{\bm{\xi},\bm{f}}\sum_{a \in \mathcal{A}^e} \gamma \cdot \xi_a \cdot \tau_a(\xi_a) + \sum_{i \in \mathcal{C}} \lambda_i \cdot d_i\\
& \qquad \qquad \quad \ s.t. \eqref{6},\eqref{7(c)},\eqref{7(d)},\eqref{7(e)}
\end{aligned}
\end{equation}

We firstly derive the KKT conditions for the upper-level variables $\bm{g},\bm{v},\bm{\theta}$ based on the Lagrangian function, i.e.,

\begin{subequations}\label{31}
\begin{align}
& \sum_{j\in\Omega_i}K_{2ij}(\lambda_i-\lambda_j+\overline{\eta_{ij}}-\overline{\eta_{ji}}-\underline{\eta_{ij}}+\underline{\eta_{ji}})+\nonumber\\
&\qquad \lambda_i \frac{\partial d_i}{\partial \theta_i}=0,\forall i \in \mathcal{N}\label{31a}\\
&\sum_{j\in\Omega_i}K_{1ij}(\lambda_i-\lambda_j+\overline{\eta_{ij}}-\overline{\eta_{ji}}-\underline{\eta_{ij}}+\underline{\eta_{ji}})+\nonumber\\
&\qquad \overline{\mu}_i-\underline{\mu}_i+\lambda_i \frac{\partial d_i}{\partial v_i}=0,\forall i \in \mathcal{N}\label{31b}\\
& c_i-\lambda_i-\underline{\tau_i}+\overline{\tau_i}+\lambda_i \frac{\partial d_i}{\partial g_i}=0,\forall i \in \mathcal{N}\label{31c}\\
& 0 \leq \bm{\underline{\tau}} \perp \bm{g} \geq 0\label{31d}\\
&0 \leq \bm{\overline{\tau}} \perp (\bm{\overline{g}}-\bm{g} )\geq 0\label{31e}\\
&0 \leq (K_{1ij}(v_i-v_j)+K_{2ij}(\theta_i-\theta_j)+f_{ij})\perp\underline{\eta_{ij}}\geq 0,\nonumber\\
&\qquad\forall i \in \mathcal{N},\forall j \in \Omega_i\label{31f}\\
&0 \leq (f_{ij}-K_{1ij}(v_i-v_j)-K_{2ij}(\theta_i-\theta_j))\perp\overline{\eta_{ij}}\geq 0,\nonumber \\
& \qquad \forall i \in \mathcal{N},\forall j \in \Omega_i\label{31g}\\
&0 \leq (v_i-\underline{v})\perp\underline{\mu_i}\geq 0,\forall i \in \mathcal{N}\label{31h}\\
&0 \leq (\overline{v}-v_i)\perp\overline{\mu_i}\geq 0,\forall i \in \mathcal{N}\label{31i}\\
& \eqref{8(c)}\label{31j}
\end{align}
\end{subequations}
where \eqref{31a}-\eqref{31c} are the stationarity conditions (which include the gradient of the upper-level variables $\bm{g},\bm{v},\bm{\theta}$ w.r.t. the lower-level variables $\bm{d}$ \cite{alesiani2023implicit}), and \eqref{31d}-\eqref{31i} are the  complementarity conditions, while \eqref{31j} ensures the primal feasibility.

As the lower-level problem is only associated with the dual variable $\lambda_i$ and therefore $d_i$ is only a function of $\lambda_i$ (as we show in \eqref{11}), the partial derivatives $\frac{\partial d_i}{\partial \theta_i},\frac{\partial d_i}{\partial v_i},\frac{\partial d_i}{\partial g_i}$ equal 0. Therefore, the stationarity constraints in \eqref{31a},\eqref{31b},\eqref{31c} can be further simplified.
With the simplified stationarity constraints, the KKT conditions of the upper-level problem are the same as the KKT conditions of IDSO problem in \eqref{8}. Obviously, the KKT conditions of the lower-level problem are the KKT conditions of ITSO in \eqref{7}. Therefore, the KKT conditions of the bilevel program \eqref{30} are in the same form of \eqref{19}, which ends the proof.
\end{document}